\documentclass[twocolumn]{aastex701}

\begin{document}


\title{\large Ly$\alpha$ Escape in JWST/NIRCam F430M-Selected H$\alpha$ Emitters at $z\simeq5.5$}
\author[orcid=0000-0003-0202-0534]{Cheng Cheng}
\affiliation{Chinese Academy of Sciences South America Center for Astronomy, National Astronomical Observatories, CAS, Beijing 100101, People’s Republic of China, chengcheng@nao.cas.cn}
\affiliation{Key Laboratory of Optical Astronomy, NAOC, 20A Datun Road, Chaoyang District, Beijing 100101, China}
\email[show]{chengcheng@nao.cas.cn}

\author[orcid=0000-0002-9634-2923]{Zhen-Ya Zheng}
\affiliation{CAS Key Laboratory for Research in Galaxies and Cosmology, Shanghai Astronomical Observatory, Shanghai 200030, China; zhengzy@shao.ac.cn} 
\email[show]{zhengzy@shao.ac.cn}

\author[orcid=0000-0002-9634-2923]{Chunyan Jiang}
\affiliation{CAS Key Laboratory for Research in Galaxies and Cosmology, Shanghai Astronomical Observatory, Shanghai 200030, China; zhengzy@shao.ac.cn} 
\email[]{zhengzy@shao.ac.cn}

\author[0000-0002-4622-6617]{Fengwu Sun}
\affiliation{Center for Astrophysics $|$ Harvard \& Smithsonian, 60 Garden St., Cambridge, MA 02138, USA}
\email[]{fengwu.sun@cfa.harvard.edu}

\author[0009-0008-9801-2224]{Edo Ibar}
\affiliation{Instituto de F\'isica y Astronom\'ia, Universidad de Valpara\'iso, Avda. Gran Breta\~na 1111, Valpara\'iso, Chile}
\affiliation{Millennium Nucleus for Galaxies (MINGAL), Chile}
\email[]{eduardo.ibar@uv.cl}

\author[0000-0002-9373-3865]{Xin Wang}
\affiliation{School of Astronomy and Space Science, University of Chinese Academy of Sciences (UCAS), Beijing 100049, China, xwang@ucas.ac.cn}
\affiliation{National Astronomical Observatories, Chinese Academy of Sciences, Beijing 100101, China}
\affiliation{Institute for Frontiers in Astronomy and Astrophysics, Beijing Normal University, Beijing 102206, China}
\email[]{xwang@ucas.ac.cn}
\author[0000-0001-7592-7714]{Haojing Yan} 
\affiliation{Department of Physics and Astronomy, University of Missouri, Columbia, MO 65211}
\email[]{yanhaojing@gmail.com }

\author[0000-0001-6763-5869]{Fang-Ting Yuan}
\affiliation{CAS Key Laboratory for Research in Galaxies and Cosmology, Shanghai Astronomical Observatory, Shanghai 200030, China; zhengzy@shao.ac.cn} 
\email[]{yuanft@shao.ac.cn}

\author[0000-0001-6511-8745]{Jia-Sheng Huang}
\affiliation{Chinese Academy of Sciences South America Center for Astronomy, National Astronomical Observatories, CAS, Beijing 100101, People’s Republic of China, chengcheng@nao.cas.cn}
\affiliation{CAS Key Laboratory of Optical Astronomy, National Astronomical Observatories, Chinese Academy of Sciences, Beijing 100101, People’s Republic of China}
\affiliation{Harvard-Smithsonian Center for Astrophysics, 60 Garden Street, Cambridge, MA 02138, USA}
\email[]{jhuang@nao.cas.cn}

\author[0000-0002-8136-8127]{Juan Molina}
\affiliation{Instituto de F\'isica y Astronom\'ia, Universidad de Valpara\'iso, Avda. Gran Breta\~na 1111, Valpara\'iso, Chile}
\affiliation{Millennium Nucleus for Galaxies (MINGAL), Chile}
\email[]{juan.molinato@uv.cl}

\author[0000-0002-0245-6365]{Malte Brinch}
\affiliation{Instituto de F\'isica y Astronom\'ia, Universidad de Valpara\'iso, Avda. Gran Breta\~na 1111, Valpara\'iso, Chile}
\affiliation{Millennium Nucleus for Galaxies (MINGAL), Chile}
\email[]{malte.brinch@uv.cl}

\begin{abstract}
We study the Ly$\alpha$ escape fraction ($f_{\rm esc}$) in an H$\alpha$-selected sample of star-forming galaxies at $z\simeq5.5$, identified via JWST/NIRCam F430M excess and covered by archival VLT/MUSE data. By anchoring the intrinsic Ly$\alpha$ production to H$\alpha$ emission, our approach provides a direct and Ly$\alpha$-unbiased probe of the escape of Ly$\alpha$ photons in galaxies with SFR $\gtrsim 0.1\,M_\odot\,{\rm yr^{-1}}$ at this epoch. Ly$\alpha$ emission is detected in 3 out of 12 galaxies covered by VLT/MUSE. Combining detections and upper limits, we place a conservative upper bound of $\langle f_{\rm esc}^{\rm Ly\alpha} \rangle < 0.32$ on the population-averaged Ly$\alpha$ escape fraction. We find that Ly$\alpha$ detections are preferentially associated with nearly dust-free systems, while no clear correlation between SFR and $f_{\rm esc}$, suggesting a stochastic picture of Ly$\alpha$ escape. Interestingly, three of the four Ly$\alpha$-detected galaxies reside within a known overdense structure, suggesting that local environment may further facilitate Ly$\alpha$ photons escape. Our H$\alpha$-selected approach establishes a general and scalable framework for probing Ly$\alpha$ escape by combining JWST medium- or narrow-band imaging with ground-based spectroscopic data, enabling systematic and less biased studies of Ly$\alpha$ visibility in typical star-forming galaxies during the post-reionization era.
\end{abstract}



\section{Introduction}

Lyman-$\alpha$ (Ly$\alpha$) emission serves as a critical probe of the high-redshift Universe, particularly for studying the epoch of reionization \citep{2014PASA...31...40D, 2016ARA&A..54..761S}. As a resonant line, Ly$\alpha$ is highly sensitive to the neutral gas content in and around galaxies, making it a powerful tracer of both galaxy formation and the intergalactic medium during cosmic reionization \citep{2015MNRAS.446..566M, 2018ApJ...857L..11M, 2021MNRAS.504.1902G}. Its intrinsic brightness and accessibility from ground-based narrow band observations have enabled the identification of large galaxy samples at $z \gtrsim 6$, underpinning our understanding of early galaxy evolution  \citep[e.g., ][]{1999ApJ...522L...9H, 2002ApJ...568L..75H, 2003ApJ...585L..97T, 2003AJ....126.2091A, 2004AJ....127..563H, 2010ApJ...725..394H, 2017ApJ...842L..22Z, 2018PASJ...70S..16K, 2018NatAs...2..962J, 2019PASP..131g4502Z, 2019ApJ...886...90H, 2021ApJ...914...79T, 2023ApJS..268...24K, 2024ApJ...971..136S}. However, the interpretation of Ly$\alpha$ emission is fundamentally complicated by resonant radiative transfer. The observed Ly$\alpha$ luminosity depends not only on the intrinsic production rate of ionizing photons, but also on the distribution, kinematics, and covering fraction of neutral gas and dust in the interstellar and circumgalactic medium, as well as attenuation by the intergalactic medium \citep{1999ApJ...518..138H, 2011ApJ...728...52L, 2014PASA...31...40D, 2018ApJ...856....2M}.

A key quantity governing the observability and interpretation of Ly$\alpha$ emission is the Ly$\alpha$ escape fraction ($f_{\mathrm{esc}}^{\mathrm{Ly}\alpha}$), defined as the fraction of intrinsically produced Ly$\alpha$ photons that escape from a galaxy and reach the observer. Measurements of $f_{\rm esc}^{\rm Ly\alpha}$ provide direct constraints on the physical conditions of the interstellar and circumgalactic medium, and are essential for calibrating Ly$\alpha$-based galaxy surveys and interpreting Ly$\alpha$ luminosity functions \citep[e.g.,][]{2022ApJ...926..230N, 2025ApJ...989...31T}. Moreover, robust constraints on $f_{\mathrm{esc}}^{\mathrm{Ly}\alpha}$ are required to disentangle internal galaxy processes from intergalactic attenuation when assessing the role of galaxies during cosmic reionization \citep[e.g.,][]{2014PASA...31...40D, 2014A&A...562A..52D, 2025ARA&A..63...45J}. 

Despite its importance, robust constraints on $f_{\mathrm{esc}}^{\mathrm{Ly}\alpha}$ at $z \gtrsim 5$ remain limited or uncertain. intrinsic Ly$\alpha$ luminosities in high-redshift Ly$\alpha$ emitter studies are commonly inferred from UV-based star formation rates \citep[e.g., ][]{2010ApJ...711..693K, 2025MNRAS.542.3125S}. However, UV continuum emission traces star formation over $\sim$100 Myr timescales, significantly longer than the $\lesssim$10 Myr timescales probed by hydrogen recombination lines \citep{2012ARA&A..50..531K}. In rapidly evolving or bursty systems, this mismatch can lead to an underestimation of the instantaneous ionizing photon production rate, particularly in low-mass galaxies \citep{2012ApJ...744...44W,2022MNRAS.511.4464A}. Such low-mass systems are thought to be important contributors to cosmic reionization \citep{2024Natur.626..975A}, and consequently, the intrinsic Ly$\alpha$ luminosity. As a result, Ly$\alpha$ escape fractions derived from UV-based star formation rates may be systematically underestimated. 

Furthermore, while Ly$\alpha$ emitter (LAE) samples have been instrumental in characterizing galaxies with prominent Ly$\alpha$ lines, they are inherently biased toward systems in which Ly$\alpha$ photons already escape efficiently, and therefore represent only a subset of the broader star-forming galaxy population \citep[e.g., $f_{\rm esc}^{\rm Ly\alpha}$ for LAEs:][]{2009A&A...506L...1A, 2024A&A...690A.302G,2024A&A...688A.106N,2025MNRAS.542.3125S}. 
Although deep observations of LAEs can probe Ly$\alpha$ escape fractions as low as $f_{\rm esc}^{\rm Ly\alpha}\sim0.01$ \citep{2024A&A...688A.106N}, such samples remain conditioned on the prior detection of Ly$\alpha$ emission. As a result, a more fundamental question remains unresolved: among typical star-forming galaxies, how frequently does Ly$\alpha$ escape, and with what efficiency? Addressing this question requires a galaxy sample selected independently of Ly$\alpha$ emission \citep{2010Natur.464..562H, 2017ApJ...835..116A}, together with a tracer of the intrinsic hydrogen recombination rate that is insensitive to resonant scattering.

JWST/NIRSpec spectroscopy provides a powerful opportunity to directly measure the Ly$\alpha$ escape fraction during the epoch of reionization, as its wavelength coverage allows simultaneous observations of Ly$\alpha$ and H$\alpha$ emission for galaxies at $z \gtrsim 5$ \citep[e.g.,][]{2024MNRAS.528.7052C, 2024A&A...688A.106N}. In principle, this enables a direct and dust-insensitive estimate of the intrinsic Ly$\alpha$ production rate based on the Balmer recombination lines under Case~B assumptions. However, practical limitations remain. In particular, aperture misalignment between the NIRSpec microshutters and galaxy centers can lead to slit losses, and the reduced spectral resolution at the blue end of the NIRSpec prism can further affect the detectability of Ly$\alpha$ emission. These aperture losses preferentially affect Ly$\alpha$ at the blue end of the prism, while H$\alpha$ at redder wavelengths benefits from higher spectral resolution. While the higher-resolution grating modes (R$\sim$1000) partially mitigate these resolution limitations and the associated aperture loss can be statistically quantified and corrected \citep{2024MNRAS.531.2701T}, spatial offsets between Ly$\alpha$ and UV continuum may still cause significant pseudo-slit losses in NIRSpec MSA observations \citep{2025MNRAS.542..128B}. 
Future NIRSpec integral field spectroscopy (IFS) observations could help avoid these aperture losses by providing spatially resolved spectroscopy. A recent study by \citet{2024ApJ...972..121J} shows that one galaxy exhibiting strong Ly$\alpha$ emission in VLT/MUSE (EW$_{\rm Ly\alpha} = 75\pm 33$ \AA) are not detected in Ly$\alpha$ with NIRSpec prism observations, highlighting the difficulty of relying solely on JWST spectroscopy to derive robust Ly$\alpha$ escape fractions for individual galaxies.

A complementary approach is therefore to combine ground-based Ly$\alpha$ measurements with JWST observations of rest-frame optical recombination lines. Ground-based facilities provide well-calibrated Ly$\alpha$ fluxes from narrow band filters or integral field spectrographs, while JWST enables sensitive measurements of H$\alpha$ emission that are largely unaffected by resonant scattering and the neutral intergalactic medium. Such a joint strategy has been demonstrated to provide improved constraints on the Ly$\alpha$ escape fraction compared to Ly$\alpha$-selected samples alone \citep[e.g., Ly$\alpha$ flux estimated from ground-based telescope, and H$\alpha$ flux from JWST data][]{2023ApJ...944L...1N, 2023MNRAS.523.5468S, 2024ApJS..272...33L, 2025arXiv250918302P}.

While recent JWST slitless spectroscopy surveys \citep[e.g., ][]{2023ApJ...953...53S, 2024ApJS..272...33L, 2025ApJ...982..153M, 2025A&A...694A.178C, 2025ApJ...987..186F, 2025MNRAS.541.1348P} have successfully assembled large statistical samples of H$\alpha$ emitters over a wide redshift range, medium-band imaging provides a complementary and robust alternative selection strategy \citep{2023ApJS..268...64W, 2026arXiv260401271L}. Medium-band photometry selects galaxies based on narrow-band flux excess and enables a spatially uniform and clean identification of emission-line sources. In crowded regions, such as those seen in galaxy clusters, slitless observations can be strongly affected by spectral overlap. In contrast, medium-band excess imaging can provide a cleaner selection criteria for high-$z$ emitters.

In this work, we characterize a H$\alpha$-bright galaxy sample selected with the medium-band filter F430M at $z \simeq 5.5$ in the Abell 2744 field (hereafter A2744), targeting typical star-forming galaxies near the end of the reionization epoch. By exploiting deep archival VLT/MUSE observations, we constrain the Ly$\alpha$ emission for each galaxy and derive Ly$\alpha$ escape fractions in a manner that is unbiased with respect to Ly$\alpha$ emitter selection. This approach allows us to investigate the physical and environmental factors governing Ly$\alpha$ escape in galaxies during the final stages of cosmic reionization. This paper is organized as follows. In Sections~2 and~3, we describe the H$\alpha$ sample selection and present the observational results. We discuss the implications of our findings in Section~4 and summarize our conclusions in Section~5. We adopt a flat $\Lambda$CDM cosmology with $H_0 = 70\,\mathrm{km\,s^{-1}\,Mpc^{-1}}$, $\Omega_{\rm M} = 0.3$, and $\Omega_{\Lambda} = 0.7$, and use AB magnitudes throughout.

\begin{figure}
    \centering
    \includegraphics[width = 0.95\linewidth]{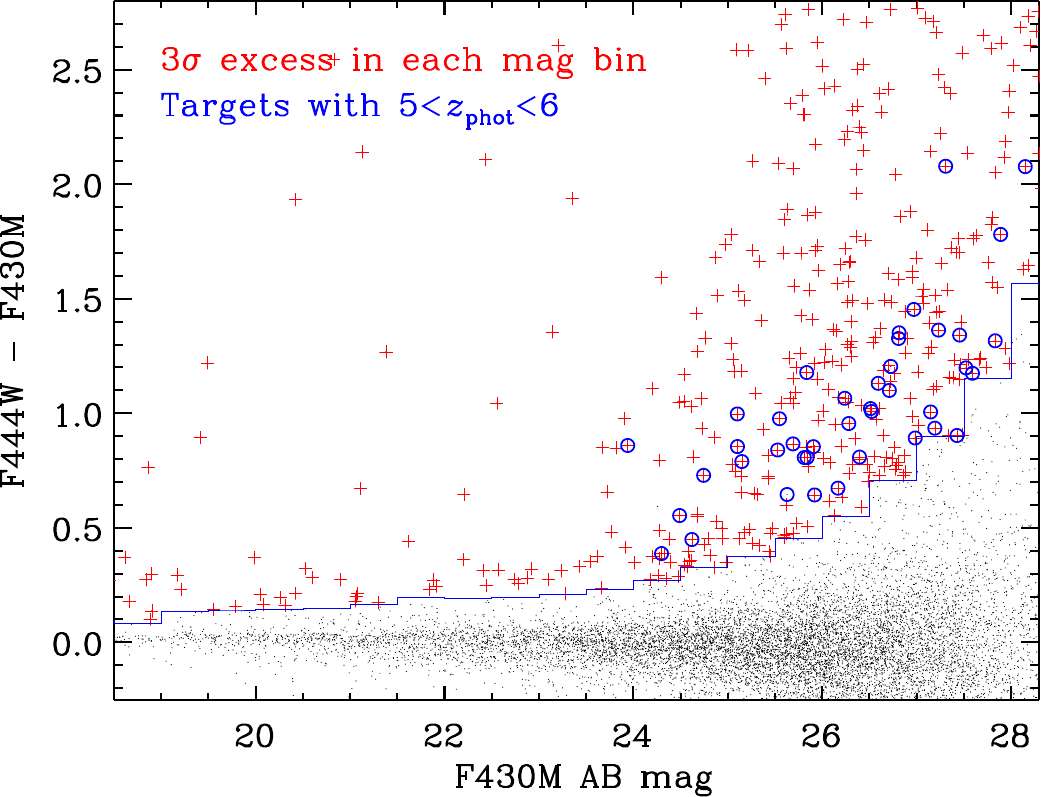}
    \includegraphics[width = 0.94\linewidth]{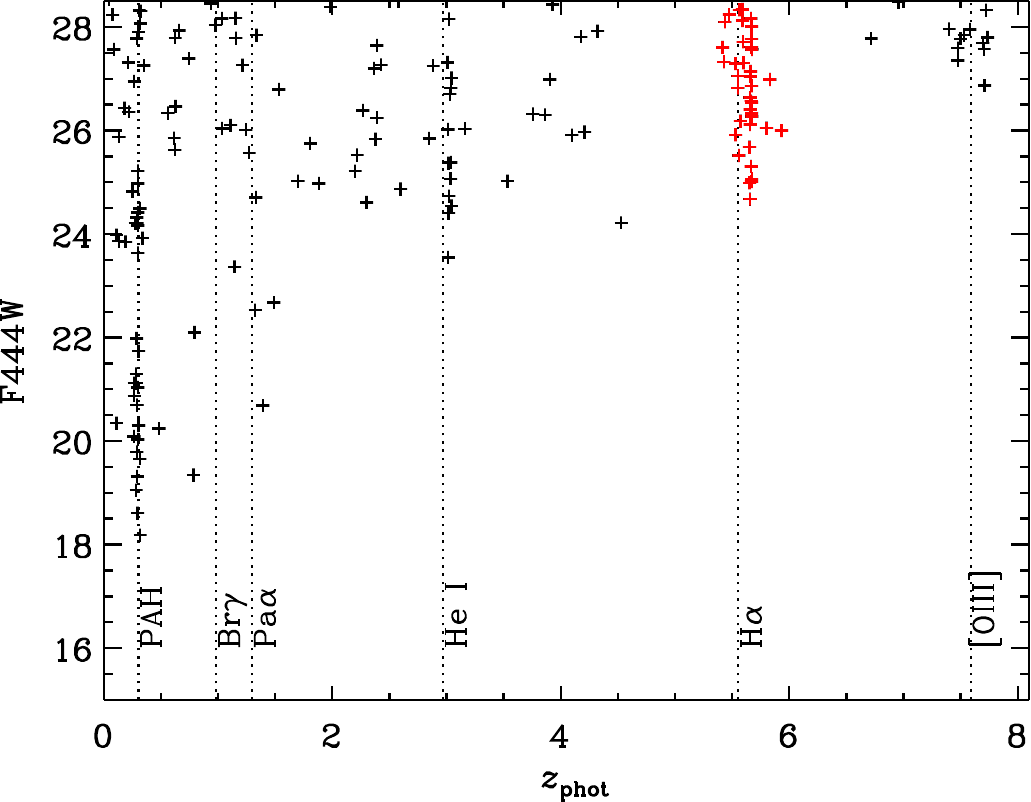}
    \caption{{\bf Upper panel}: F444W$-$F430M color vs. F430M magnitude for the F430M image-selected galaxies. Red plus signs mark sources with a $3\sigma$ excess in the F430M$-$F444W color, and the solid line shows the $3\sigma$ detection limit evaluated in 0.5~mag bins. Blue circles highlight the F430M emitters with $5<z_{\rm phot}<6$.
    {\bf Lower panel}: F444W magnitude vs.\ photometric redshift for the F430M excess sample. Dotted lines indicate the redshifts at which emission lines shift into the F430M bandpass. Targets with $5<z_{\rm phot}<6$ are highlighted in red.
    }
    \label{preselection}
\end{figure}
 
\begin{figure}
    \centering
    \includegraphics[width = 0.95\linewidth]{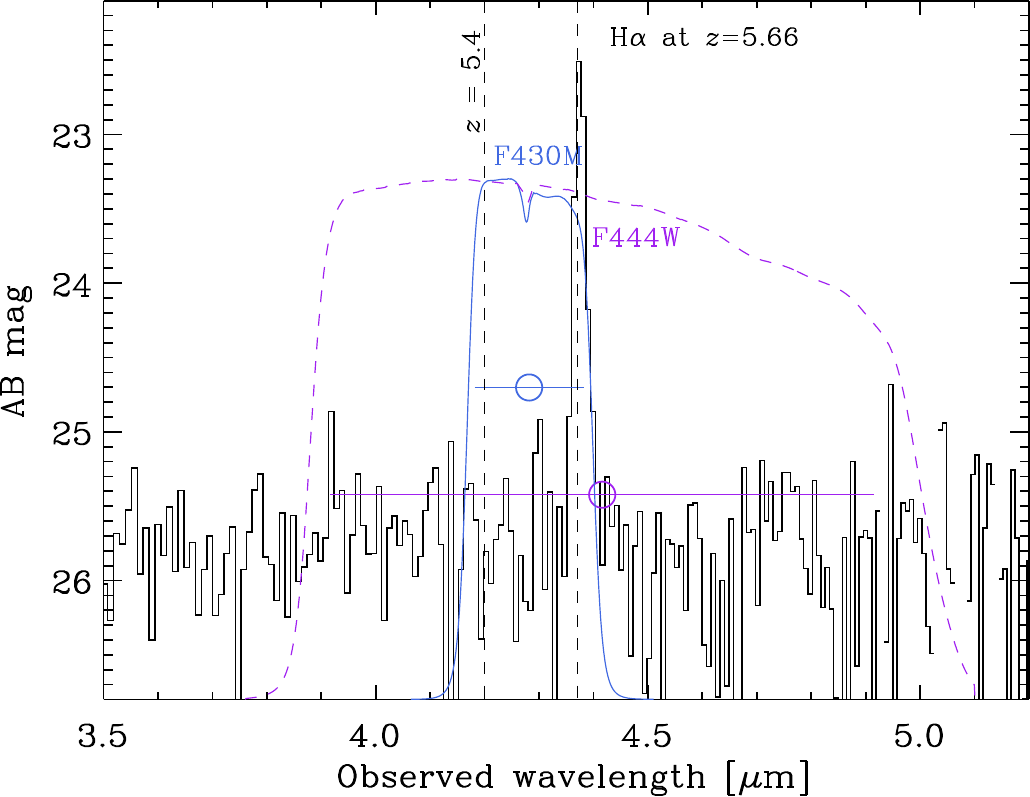}
    \includegraphics[width = 0.95\linewidth]{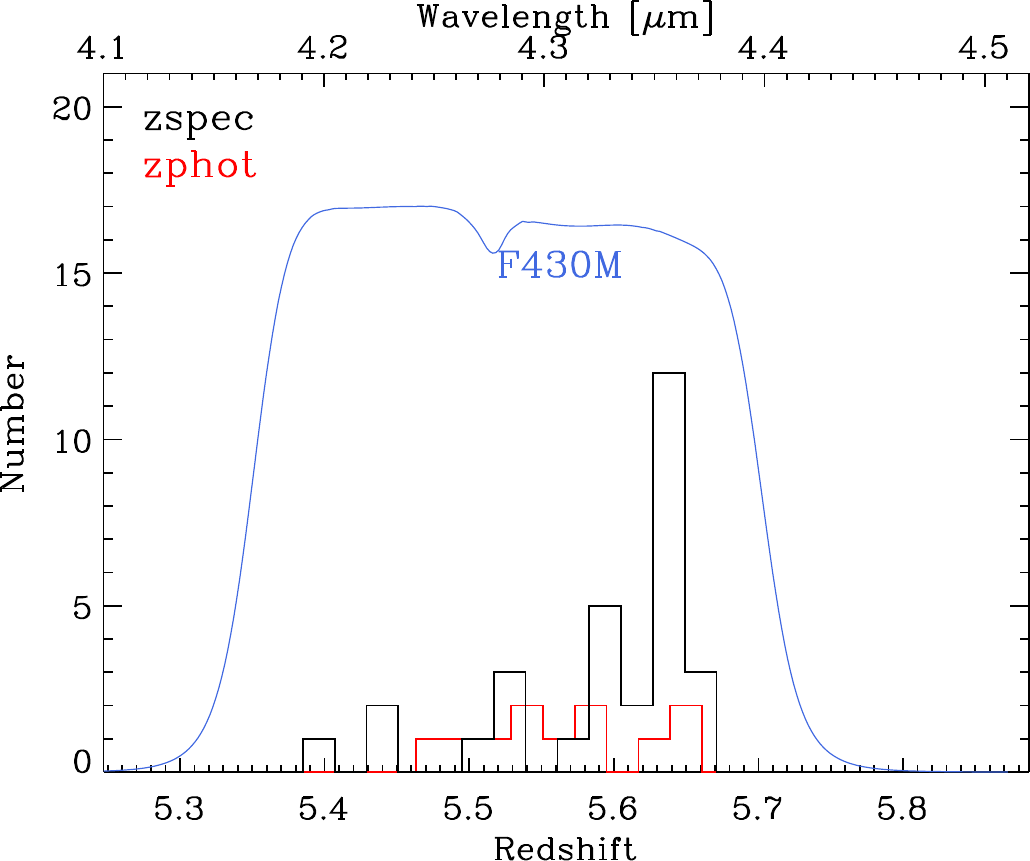}
    \caption{{\bf Upper panel}: Response curves for the JWST/NIRCam filter F430M, F444W. We show the NIRSpec spectrum for one H$\alpha$ emitter ID 27262, which is downloaded from DAWN JWST Archive \citep{2024Sci...384..890H, 2025A&A...697A.189D}. The emission line is the H$\alpha$ at $z = 5.66$. The dashed lines show the redshifts for H$\alpha$ emission at redshift $z = 5.66$, as well as the redshift 5.4, which are the range of the available redshifts for H$\alpha$ emitters selected from the F430M filter. The open circles are the flux measured from the spectrum.
    {\bf Bottom panel}: The histogram of the redshifts for the selected H$\alpha$ emitter (red for photometric redshifts and black for the spectroscopic redshifts). The corresponding wavelength is indicated on the upper x‑axis. The blue curve is the response curve for F430M filter.
    }
    \label{targetselection}
\end{figure}

\begin{figure*}
    \centering
    \includegraphics[width=0.99\linewidth]{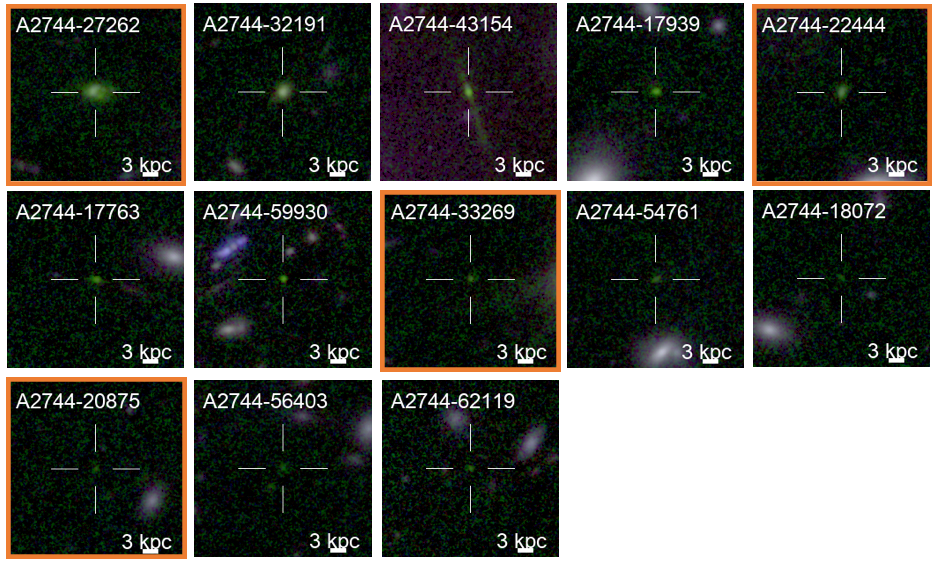}
    \caption{JWST/NIRCam false-color stamps (F360M, F430M, F444W assigned to blue, green, red) of the H$\alpha$ sample in this work. Orange boxes highlight the four targets with Ly$\alpha$ emission detected in the MUSE data cube. Each stamp is $6'' \times 6''$. The white bar in the lower right indicates $0.5''$ ($\sim$3 kpc at $z = 5.5$).
    }
    \label{stamp}
\end{figure*}

\section{Sample selection and reduction}

\subsection{H$\alpha$ emitter at z $\sim$ 5.5 selection}

Following the selection strategy of \citet{2025ApJS..279...43C}, we identify a sample of F430M line emitters by requiring a $3\sigma$ excess in the F430M$-$F444W color. We additionally impose a flux cut of F430M$<28.5$, corresponding to the $\sim5\sigma$ depth of the F430M image measured in a circular aperture of radius $0.13''$ \citep{2024arXiv241001874N}, which ensures robust detections of F430M excess emitters. 
The F430M excess targets are shown in the upper panel of Figure \ref{preselection}. Then we use the following selection steps to narrow down the F430M emitters to H$\alpha$ emitters at $z\sim 5.5$.

\begin{itemize}
\item {\bf Broadband catalog cross-match} We cross-match the selected F430M excess emitters with the SUPER photometric catalog from the UNCOVER project \citep{2024ApJ...976..101S, 2024ApJ...974...92B} to obtain the multi-band photometry and photometric redshift estimates. 
This step also excludes spurious F430M excess detections caused by stellar diffraction spikes, cosmic rays, or image artifacts, which lack counterparts in the deep broadband images.
The UNCOVER catalog provides homogeneous photometry from UV to the F480M band, enabling tight constraints on photometric redshifts for high-redshift galaxies.
We show the redshift and F444W magnitude distribution in the lower panel of Figure \ref{preselection}.
We can see that
the resulting redshift distribution reveals several distinct groups, corresponding to different emission-line populations whose prominent lines fall within the F430M bandpass, including PAH emission at low redshift, Pa$\alpha$ and He\,\textsc{i} 10830\AA\ at intermediate redshift, and H$\alpha$ emission at $z \simeq 5.5$.

\item {\bf Photometric redshift selection} 
We select emitters with $5 < z_{\rm phot} < 6$. For sources at $z \simeq 5.5$, the redshifts are jointly constrained by the presence of strong emission lines entering the medium-band filters and by the Ly$\alpha$ break sampled by the F090W bands. As a result, the identification of H$\alpha$ emitters with photometric redshifts around $z_{\rm phot} \sim 5.5$ is highly robust.
\end{itemize}

This selection yields 41 H$\alpha$-bright galaxies within an area of $\sim30~\mathrm{arcmin}^2$, which constitute the parent sample of this study.
Because the F430M is a medium-band filter with $\Delta\lambda \simeq 2315 $\AA, the line flux is diluted by a factor of $\sim$ 20 relative to the continuum, so this selection naturally favors sources with high rest-frame equivalent widths (EW$_0 \gtrsim 200–1000 $\AA) without requiring an explicit EW cut.
A detailed description of the filter selection, target identification can be found in \citet{2025ApJS..279...43C}. In Figure~\ref{targetselection}, we show the spectrum of one H$\alpha$ emitter ID~27262 from the DAWN JWST Archive \citep{2024Sci...384..890H, 2025A&A...697A.189D}, illustrating that the wavelength coverage of the F430M filter selects H$\alpha$ emitters over the redshift range $5.4 \lesssim z \lesssim 5.66$. We crossmatch our sample with the spec-z catalog released by All the Little Things \citep[ALT, ][]{2024arXiv241001874N}, which is a JWST NIRCam grism survey covering $\sim$30 arcmin$^2$ in the A2744 field, designed to obtain deep ($\sim$7--27 hr) slitless spectroscopy at 3--4 $\mu$m and measure spectroscopic redshifts for faint galaxies down to $m_{\rm F356W} \sim 28$ AB mag. We find that 30 of the F430M-selected H$\alpha$ emitters have spec-z. We show the histogram of the photometric and spectroscopic redshift sample in bottom panel of Figure \ref{targetselection}.

\begin{figure*}
    \centering
    \includegraphics[width=0.98\linewidth]{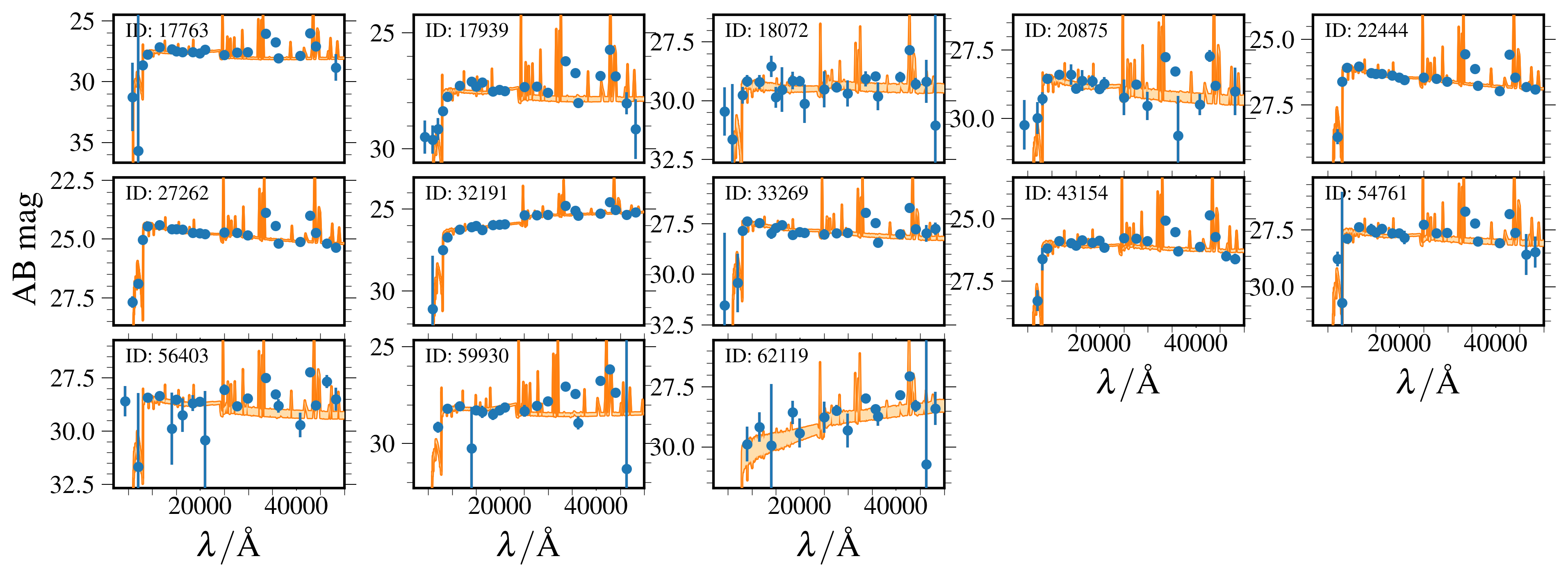}
    \includegraphics[width=0.98\linewidth]{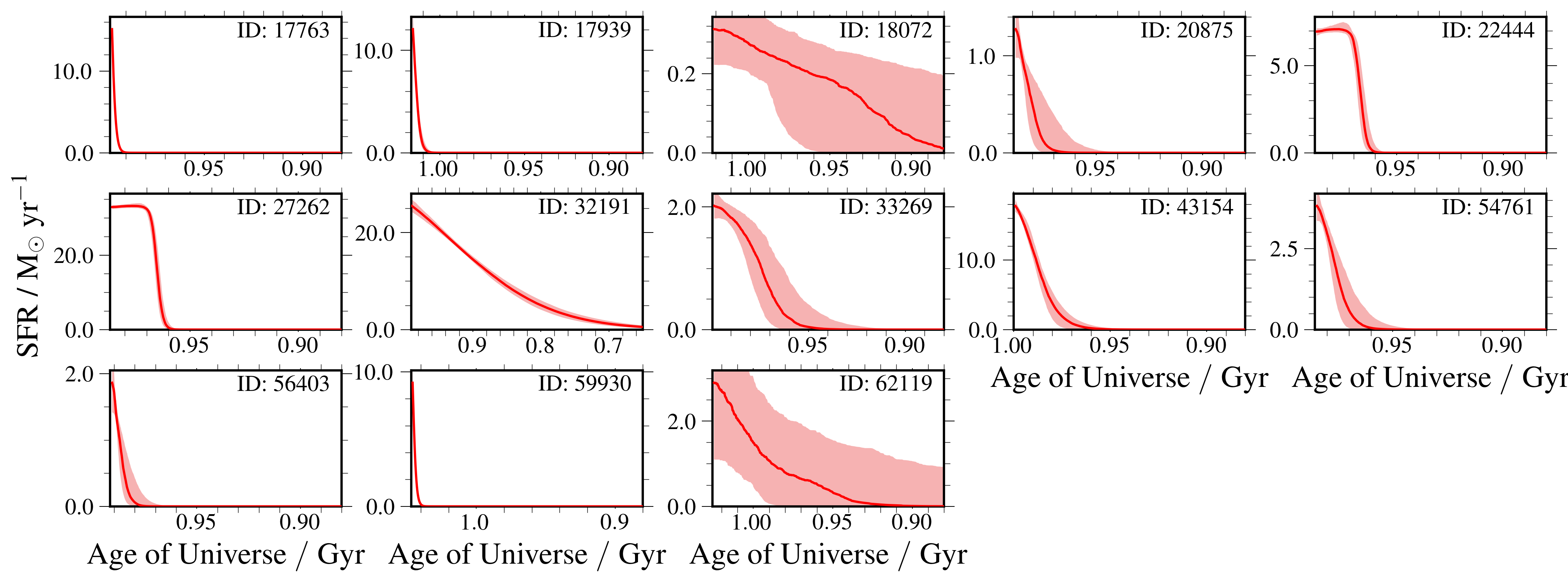}
    \caption{{\bf upper panel}: SEDs of the H$\alpha$ selected sample. The orange lines are the model spectra, and the blue dots show the photometric data from the SED released by UNCOVER \citep{2024ApJS..270....7W}. {\bf Lower panel}: Star formation histories (SFHs) of the H$\alpha$-selected sample derived with \textsc{Bagpipes}, assuming a double power-law parameterization. The inferred SFHs indicate enhanced recent star formation relative to the past-averaged activity.}
    \label{sed}
\end{figure*}

\subsection{VLT/MUSE spectroscopy and the final H$\alpha$ sample}

We download all the released VLT/MUSE cubes from ESO archive data website \footnote{\url{https://archive.eso.org/scienceportal/home?pos=3.5933333,-30.393633&r=0.15&dp_type=CUBE&ins_id=MUSE}}, which includes several previous projects that covered the center A2744 field and the outskirts of A2744. Typical exposure time for each cube is about 1 to 3 hours. The availability of MUSE coverage is determined purely by the footprint of the archival observations and is independent of the Ly$\alpha$ properties of the galaxies. As a result, a subsample of 13 H$\alpha$ emitters have MUSE coverage. These 13 Ly$\alpha$-covered and H$\alpha$-detected galaxies at $z\sim 5.5$ constitute the main sample analyzed in this work. We show the F360M, F430M, F444W fake-color images in Figure \ref{stamp}.

We cross-match our sample with published strong-lensing catalogs \citep{2023A&A...670A..60B, 2024arXiv241001874N, 2025ApJ...982...51P} and find that two sources \citep[ID 22444 and 27262, which have the IDs 18.1a and 18.1c in][]{2023A&A...670A..60B}
are identified as multiple images of the same system. Treating them as independent objects does not affect our statistical results: since $f_{\rm esc}^{\rm Ly\alpha}$ is a flux ratio independent of the absolute magnification, treating the two images separately effectively gives each line of sight equal weight, which is conservative for our statistical analysis (see Table~\ref{tab1} for a comparison of their intrinsic properties). We note that the high magnification of 27262 may introduce differential lensing effects \citep{2012ApJ...761...20H}; see the note to Table~\ref{tab1}.
We therefore retain both entries in the analysis for consistency with the photometric catalog.

\subsection{Spectral energy distribution fitting}\label{sedfitting}

We fit the spectral energy distributions (SEDs) of the H$\alpha$ sample by \textsc{Bagpipes} \citep{2018MNRAS.480.4379C, 2019MNRAS.490..417C} with the PSF-matched photometric catalog from UNCOVER \citep{2024ApJS..270....7W}. The photometry bands are F435W, F606W, F814W from HST and F070W, F090W, F115W, F140M, F150W, F162M, F182M, F200W, F210M, F250M, F277W, F300M, F335M, F356W, F360M, F410M, F430M, F444W, F460M, F480M from JWST. The parameters for \textsc{Bagpipes} include the Kroupa initial mass function \citep{2002MNRAS.336.1188K}, covering a stellar mass range of 0.1--100 $M_\odot$,
ionization parameter $\log U = -3$, 
( where $U$ is the dimensionless ionization parameter defined as the ratio of ionizing photon density to hydrogen density at the inner face of the nebula
), and the attenuation curve by \citet{calzettiDustContentOpacity2000}. We adopt a double power-law star formation history (SFH), as implemented in \textsc{Bagpipes}, which provides a flexible yet compact description of star formation histories at high redshift \citep{2018MNRAS.480.4379C}. Spectroscopic redshifts are adopted in {\sc Bagpipes} when available, and the rest targets were set at a photometric redshift with a range of [5, 6]. The SED and SFH fitting results are shown in Figure \ref{sed}. The sharp Ly$\alpha$ break and the strong [OIII] and H$\alpha$ emission lines are clearly captured by the SEDs, which are helpful in constraining the redshifts and stellar population. 

\subsection{H$\alpha$ flux Measurements}

To estimate the H$\alpha$ emission-line flux, a nearby line-free band is commonly used to characterize the underlying continuum. From Figure \ref{sed} we can see that the F444W bands still affected by the emission line. Therefore, we linearly interpolate the continuum for F430M based on the best fitted templates at the rest-frame wavelength ranges of [6430, 6470]\AA\ and [6800, 6830]\AA, where there are no bright emission lines, and close to the H$\alpha$ emission lines. We measure the interpolated flux from best fitted SED and take this as $\rm F430M_{cont}$. Then we estimate the H$\alpha$ flux by:
\begin{equation}\label{f430mexcess}
    F_{\rm H\alpha} = \Delta {\rm F430M} \left(f_{\rm F430M} - f_{\rm F430M_{\rm cont}}\right),
\end{equation}
where $F_{\rm H\alpha}$ is the line flux in units of $\rm erg\, s^{-1}\, cm^{-2}$, $f_{\rm F430M}$ and $f_{\rm F430M_{\rm cont}}$ are the flux densities in units of $\rm erg\, s^{-1}\, cm^{-2}\, \AA^{-1}$. The $\Delta {\rm F430M} = 2315.31$\AA\ is the full width at half maximum width of the F430M filter. As shown in the bottom panel of Figure~\ref{targetselection}, the spectroscopic redshifts indicate that the H$\alpha$ line of our targets falls within the flat transmission region of the F430M filter, where the throughput varies only weakly with wavelength. The conversion from F430M flux excess to H$\alpha$ flux is therefore largely insensitive to small redshift differences across the sample, and no additional correction for the filter transmission curve is applied. We also neglect the contamination from [NII] and [SII] in the medium-band flux. At these redshifts, JWST/NIRSpec spectroscopy has shown that low-mass star-forming galaxies typically have low metallicities and very weak nitrogen and sulfur emission lines, with [NII] often undetected and [SII] relatively faint compared to H$\alpha$ \citep[e.g.,][]{2023ApJ...950L...1S}.

The observed H$\alpha$ fluxes are converted to luminosities using the redshifts of the galaxies. A comparison with the traditional broad/narrow-band excess method is presented in Appendix \ref{Haflux} to validate our interpolation-based approach.

To correct the observed H$\alpha$ luminosity for dust attenuation, we adopt the stellar continuum extinction $A_V$ derived from SED fitting using \textsc{Bagpipes}. The extinction at the wavelength of H$\alpha$ is computed following the attenuation law of \citet{2000ApJ...533..682C}:
\begin{equation}
A_{\rm H\alpha} = \frac{k_{\rm H\alpha}}{f\, k_V} \, A_V,
\end{equation}
where $k_{\rm H\alpha}/k_V = 0.82$ based on the Calzetti curve value at 6563~\AA. Here $f = E(B-V)_{\rm star}/E(B-V)_{\rm neb}$ represents the ratio between stellar and nebular reddening. We adopt $f=0.44$ following \citet{2000ApJ...533..682C}, consistent with recent measurements for high-redshift star-forming galaxies \citep[e.g.,][]{2025ApJ...980...12S, 2026ApJ...997..319T}. 
Given the very low dust attenuation in our sample, adopting different values of $f$ would not significantly affect the inferred H$\alpha$ luminosities. The dust-corrected H$\alpha$ luminosity is then
\begin{equation}
L_{\rm H\alpha}^{\rm int} = L_{\rm H\alpha}^{\rm obs} \times 10^{0.4 A_{\rm H\alpha}}.
\end{equation}

\begin{figure}
    \centering
    \includegraphics[width=0.95\linewidth]{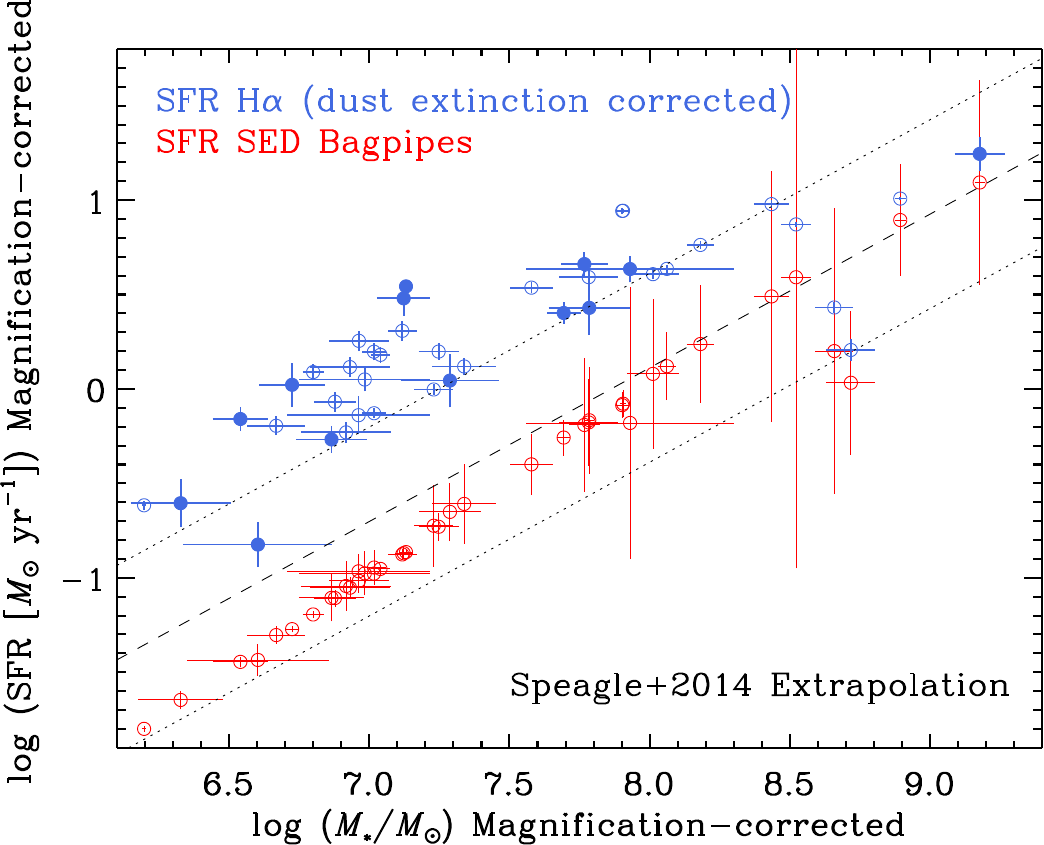}
    \caption{Star formation rate of the H$\alpha$ emitter sample. Star formation rates estimated from SED fitting with \textsc{Bagpipes} (SFR$_{\rm SED}$) are shown as red circles, while those derived from H$\alpha$ emission (SFR$_{\rm H\alpha}$) are shown as blue circles. Sources covered by MUSE observations are highlighted with solid symbols. The star-forming main sequence at $z\sim5.5$ from \citet{2014ApJS..214...15S} is shown in dotted and dashed lines for comparison, which is consistent with the SFR$_{\rm SED}$. The systematically elevated SFR$_{\rm H\alpha}$ relative to SFR$_{\rm SED}$ is consistent with bursty star formation histories in low-mass galaxies \citep[e.g., ][]{2026arXiv260120930B}. Stellar masses and star formation rates shown here have been corrected for gravitational lensing magnification.
    }
    \label{ms}
\end{figure}

To estimate the intrinsic stellar mass and star formation rate, we adopt lensing magnification factors from the publicly available LENSTOOL models\footnote{\url{https://archive.stsci.edu/prepds/frontier/lensmodels/webtool/magnif.html}} \citep{2011ascl.soft02004K, 2017ApJ...837...97L, 2022ApJ...928...87F}, as provided by the Frontier Fields Lens Modeling Comparison Project \citep{2017MNRAS.472.3177M}. For each source, we adopt the median magnification from the different lens models as the representative value, and use the robust standard deviation of the model predictions as the corresponding uncertainty. The magnifications range from $\mu \sim 2$ to 11 (listed in Table~\ref{tab1}). We show the stellar mass and star formation rate in Figure \ref{ms}. The star formation rate derived from H$\alpha$ emission (SFR$_{\rm H\alpha}$) is estimated following the calibration of \citet{2012ARA&A..50..531K}, assuming a Chabrier initial mass function. We find that SFR$_{\rm H\alpha}$ is systematically higher than the SFR inferred from SED fitting with \textsc{Bagpipes}, suggesting enhanced recent star formation relative to the longer-timescale SED-based estimates, consistent with the SED-inferred star formation histories shown in Figure~\ref{sed}. 
This is partly by construction: Bagpipes defines SFR as the average over the past 100~Myr (\texttt{sfr\_timescale = 10$^8$~yr}), so even when the model correctly captures the H$\alpha$ excess in F430M, the resulting SFR reflects the 100~Myr-averaged activity rather than the instantaneous rate traced by SFR$_{\rm H\alpha}$.
Our H$\alpha$-selected sample predominantly probes galaxies with SFR $\gtrsim 0.1\,M_\odot\,{\rm yr^{-1}}$, which is lower than the typical SFR limits ($\sim 1\,M_\odot\,{\rm yr^{-1}}$) reached by JWST slitless spectroscopy surveys at redshift $\sim$ 5 \citep[e.g., ][]{2023ApJ...953...53S, 2024ApJS..272...33L, 2025ApJ...982..153M, 2025A&A...694A.178C, 2025ApJ...987..186F}, or narrow-band selected targets \citep{2025MNRAS.541.1348P}, owing to the combined effects of gravitational lensing and medium-band selection.

\begin{figure}
    \centering
    \includegraphics[width=0.94\linewidth]{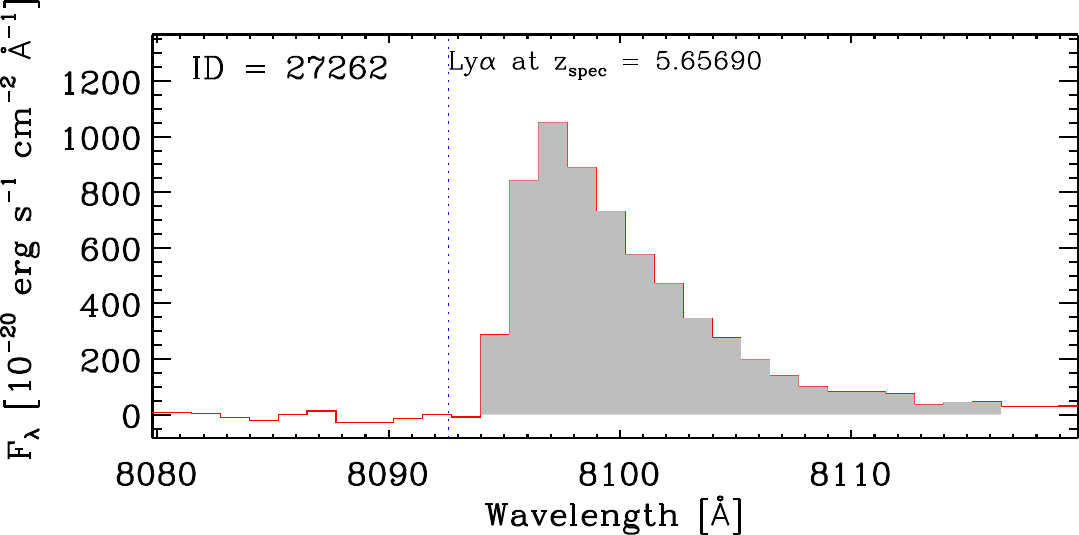}
    \includegraphics[width=0.94\linewidth]{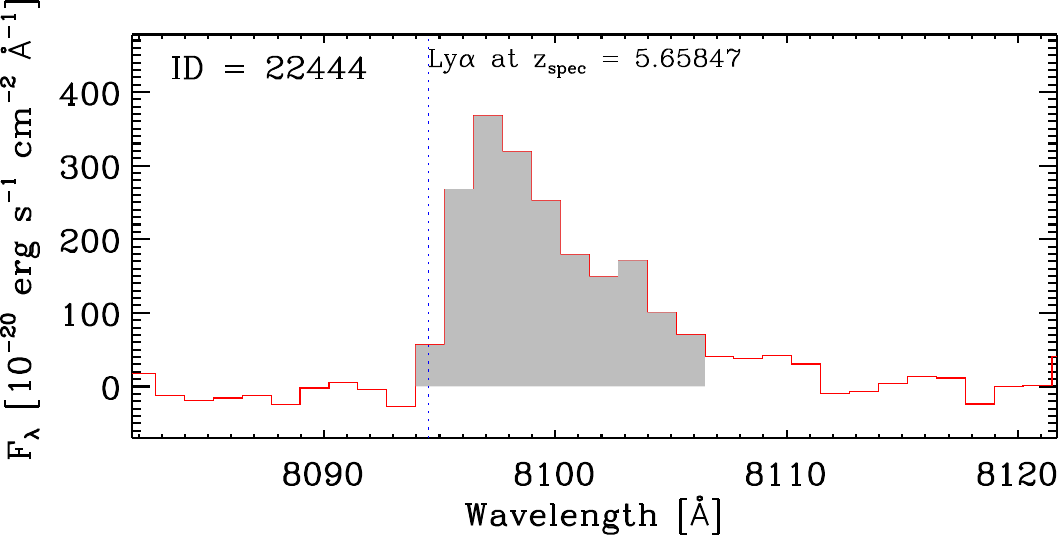}
    \includegraphics[width=0.94\linewidth]{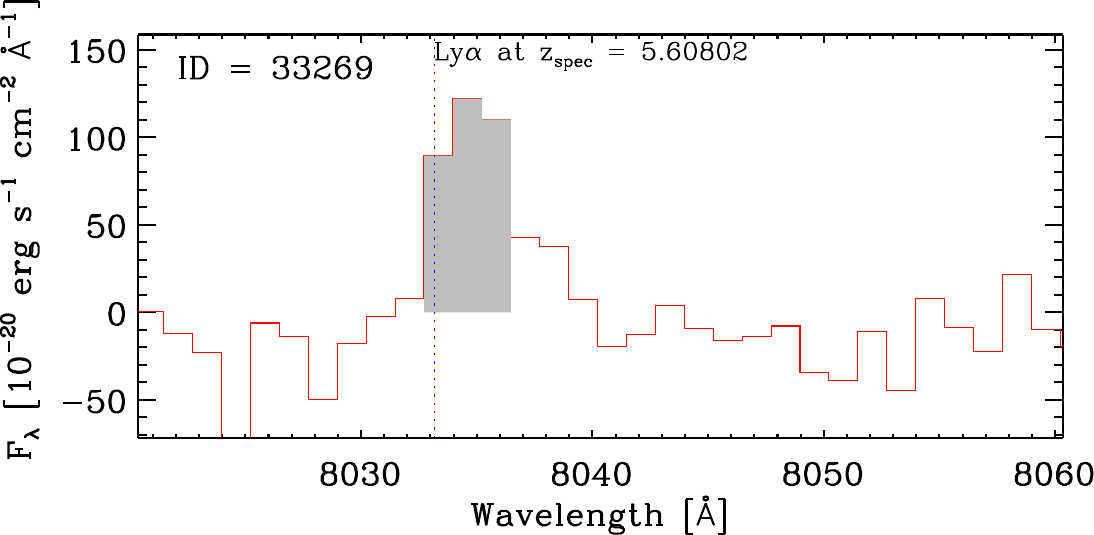}
    \includegraphics[width=0.94\linewidth]{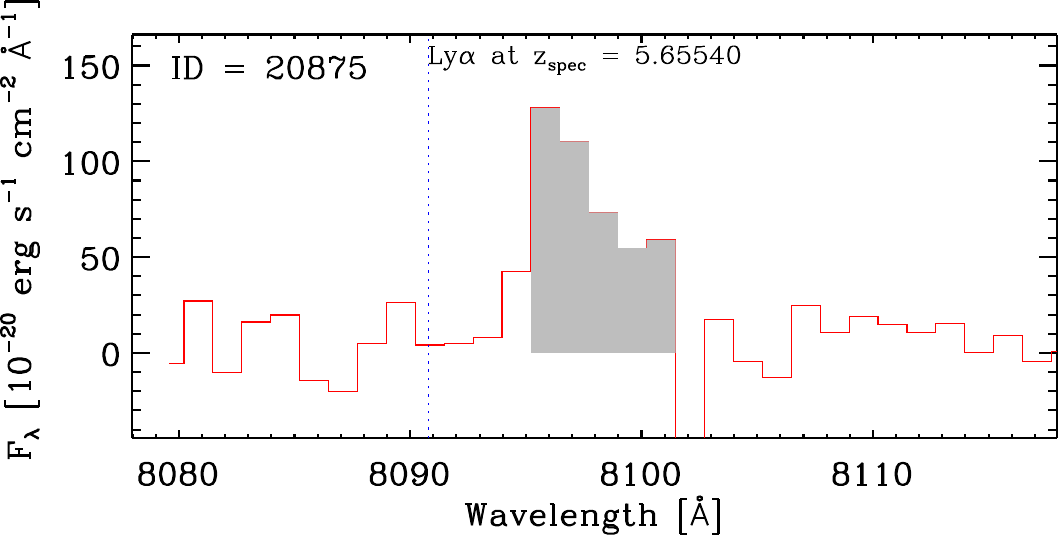}
    \caption{The VLT/MUSE spectra of the four Ly$\alpha$ detected targets. The dotted lines show the wavelength of $1216\times (1+z_{\rm spec})$ \AA. The shaded regions indicate the spectral channels with flux densities exceeding $3\sigma$ of the rms noise, which are integrated to obtain the total Ly$\alpha$ flux.}
    \label{lya_spec}
\end{figure}

\begin{figure*}
    \centering
    \includegraphics[width=0.95\linewidth]{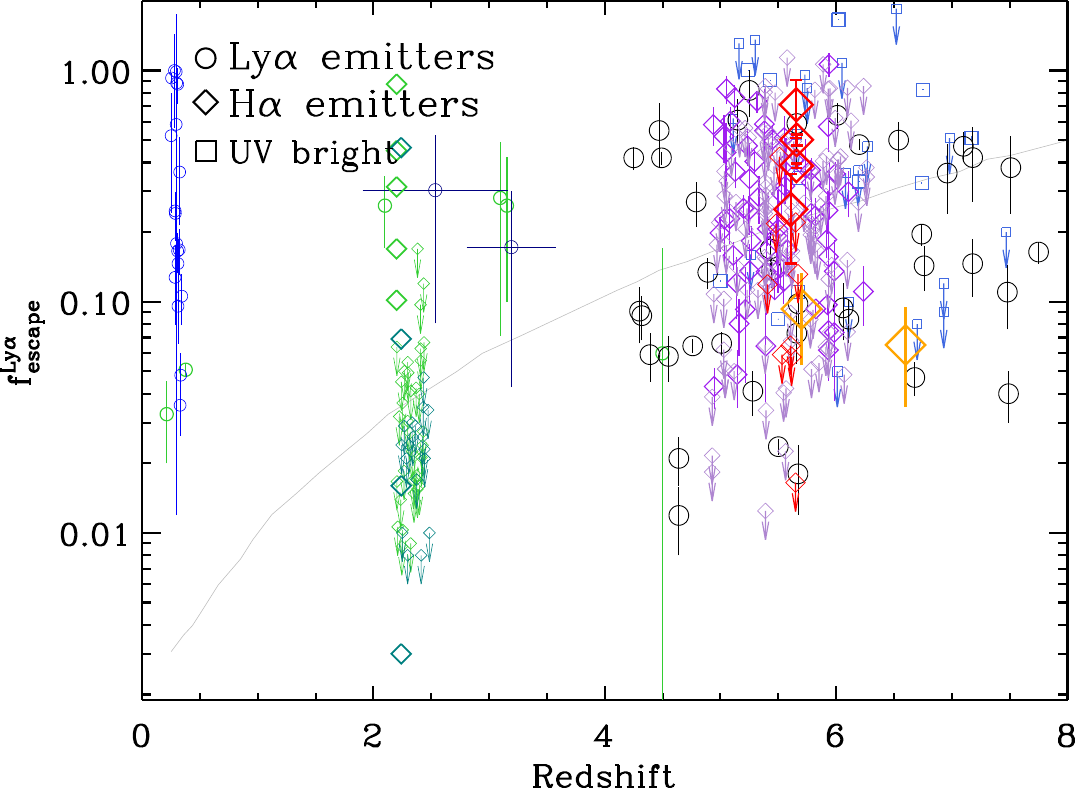}
    \caption{The Ly$\alpha$ escape fraction as a function of redshift. Our results are shown in red diamonds or red upper limits. We list the results from literature, including the Ly$\alpha$ emitters (open circles) from 
    \citet[][navy blue]{2011ApJ...736...31B},
    \citet[][lime green]{2012ApJ...746...28Z},
    \citet[][blue]{2009A&A...506L...1A},
    \citet[][black]{2024A&A...688A.106N},
    the UV selected galaxy sample (blue open squares) from \citet{2024MNRAS.528.7052C}, and the H$\alpha$ emitter (open diamonds) results of 
    \citet[][in green]{2010Natur.464..562H},
    \citet[][in dark green]{2017ApJ...835..116A},
    \citet[][in purple]{2024ApJS..272...33L},
    \citet[][in orange]{2023ApJ...953...53S},
    and this work (open diamond in red). We add a random value about 0.1 to the redshifts to avoid the overlap. To avoid overplotting of data points from different studies, we have applied a small random offset ($\sim 0.1$) to the sample of \citet{2009A&A...506L...1A, 2010Natur.464..562H} and \citet{2017ApJ...835..116A}. This visual adjustment is made solely for display purposes.
    The gray solid line is the redshift evolution trend of $f_{\rm esc}^{\rm Ly\alpha}$ taken from \citet{2011ApJ...730....8H}.
    }
    \label{fesc}
\end{figure*}

\subsection{Ly$\alpha$ Detections and Upper Limits}

We search for Ly$\alpha$ emission in the MUSE integral-field spectroscopic data at the positions of the H$\alpha$ emitters. For each source, we extract spectra from the MUSE datacube using an aperture with $1.5''$ diameter, and set the ring region with diameter between 3 and 7 arcsec as background region. The spectrum extraction diameter corresponds to about 9 kpc for the galaxies, and would include almost all the flux. Ly$\alpha$ emission is considered detected when a significant emission feature is present at the expected wavelength based on the redshift.

Ly$\alpha$ emission is clearly detected in four out of the 13 H$\alpha$ emitters with MUSE coverage (Figure~\ref{lya_spec}). Given the asymmetric Ly$\alpha$ line profiles, we adopt a direct channel-integration approach rather than parametric line fitting. Specifically, the Ly$\alpha$ flux is measured by integrating over the spectral channels where the emission exceeds $3\sigma$ of the rms noise. We list the results in Table \ref{tab1}. In addition to the integrated flux, we inspect the Ly$\alpha$ line shape to assess potential asymmetries or velocity offsets, which may provide qualitative information on resonant scattering and gas kinematics.

For the remaining nine galaxies, no significant Ly$\alpha$ signal is detected. For these non-detections, we derive the 3$\sigma$ upper limits on the Ly$\alpha$ flux based on the local noise properties of the MUSE data at the wavelength range of [1207, 1213]\AA\ and [1220, 1225]\AA. We assume a representative line width of 200 km/s \citep{2024A&A...690A.343P} when converting the noise level into a flux limit. These upper limits are incorporated into our analysis to place constraints on the Ly$\alpha$ escape fraction across the full H$\alpha$-selected sample. The combination of detections and non-detections enables a statistical assessment of the Ly$\alpha$ escape fraction in a H$\alpha$-selected galaxy sample.

We note that resonant scattering can redistribute Ly$\alpha$ photons over large spatial scales, potentially lowering the surface brightness below the detection threshold even when the total Ly$\alpha$ luminosity is non-zero \citep{2014PASA...31...40D, 2017A&A...608A...8L, 2018Natur.562..229W, 2024MNRAS.531.2701T}.

\section{Results of Ly$\alpha$ Escape Fraction}

The Ly$\alpha$ escape fraction ($f_{\mathrm{esc}}^{\mathrm{Ly}\alpha}$) is defined as the ratio of the observed Ly$\alpha$ luminosity to the intrinsic Ly$\alpha$ luminosity produced by hydrogen recombination. Under Case B recombination, the intrinsic Ly$\alpha$ luminosity relates to the dust-corrected intrinsic H$\alpha$ luminosity as $L_{\rm Ly\alpha}^{\rm int} =8.7\times L_{\rm H\alpha}^{\rm int}$ \citep[e.g., ][]{2003adu..book.....D}, since H$\alpha$ emission is largely unaffected by resonant scattering and traces the total hydrogen recombination rate. We adopt Case B (Ly$\alpha$/H$\alpha$ = 8.7) for consistency with previous studies. In the optically thin limit (Case A), the intrinsic Ly$\alpha$/H$\alpha$ ratio is $\sim$11 at $T=10{,}000$ K \citep[][Table~4.1]{2006agna.book.....O}, which would lower the derived $f_{\rm esc}^{\rm Ly\alpha}$ by $\sim$24\%. A fully Case A regime is unlikely for typical star-forming nebulae, as it requires the gas to be optically
thin to all Lyman-series photons. Thus, the Ly$\alpha$ escape fraction for each galaxy can be written as

\begin{equation}
f_{\mathrm{esc}}^{\mathrm{Ly}\alpha} =
\frac{L_{\mathrm{Ly}\alpha}^{\mathrm{obs}}}
{8.7 \times L_{\mathrm{H}\alpha}^{\mathrm{int}}}.
\end{equation}

This provides a direct comparison between observed Ly$\alpha$ and the intrinsic production inferred from H$\alpha$, independent of Ly$\alpha$-based selection. For the non-detected sources, the derived values therefore represent upper limits on $f_{\mathrm{esc}}^{\mathrm{Ly}\alpha}$, making our estimates of the average escape fraction conservative.

\begin{figure*}
    \centering
    \includegraphics[width=0.95\linewidth]{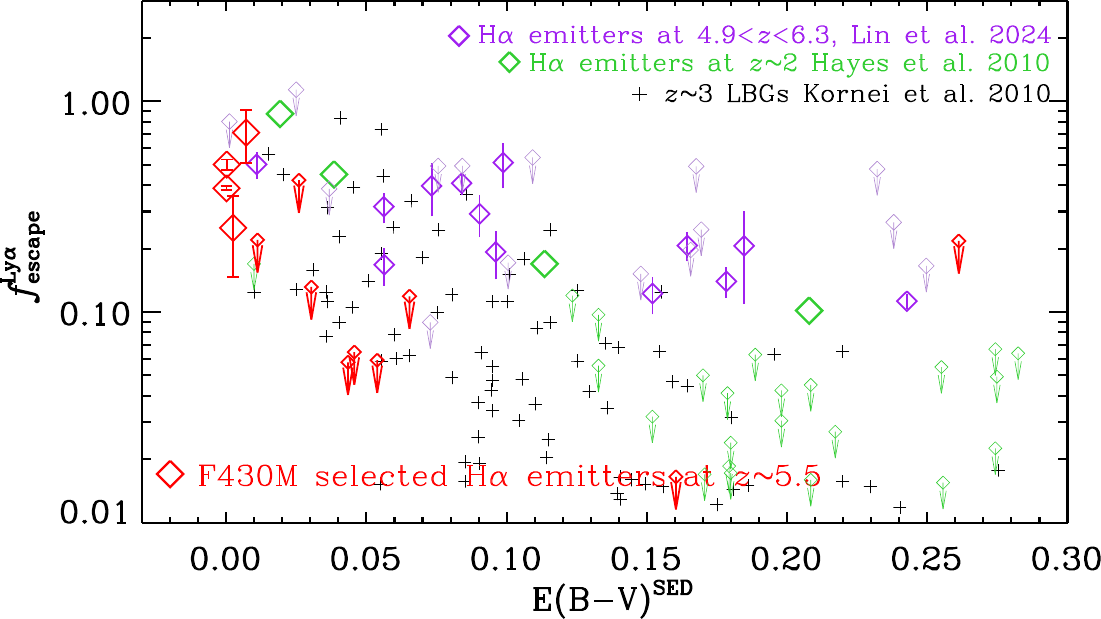}
    \caption{Distribution of the $f_{\rm esc}^{\rm Ly\alpha}$ with the E(B-V) from SED fitting (red diamonds or upper limits), comparing with the H$\alpha$ emitters at $z\simeq 2.2$ \citep{2010Natur.464..562H}, and the lyman break galaxy sample at $z\simeq 3$ \citep{2010ApJ...711..693K}. The E(B-V) is estimated from $A_V$/3.1.
    }
    \label{AvLHa}
\end{figure*}

We estimate the Ly$\alpha$ escape fraction for each of the 13 H$\alpha$ emitters with MUSE coverage, and show the results in Figure \ref{fesc}. We also compare our measurements with previous constraints on the Ly$\alpha$ escape fraction from UV-selected \citep{2024MNRAS.528.7052C}, Ly$\alpha$-selected \citep{2011ApJ...736...31B, 2012ApJ...746...28Z, 2009A&A...506L...1A, 2024A&A...688A.106N}, and H$\alpha$-selected galaxy samples \citep{2010Natur.464..562H, 2017ApJ...835..116A, 2024ApJS..272...33L} across a wide redshift range. At $z \gtrsim 5$, recent JWST-enabled studies extend the $f_{\rm esc}^{\rm Ly\alpha}$ measurements to $\lesssim 0.01$ \citep{2024ApJS..272...33L, 2024A&A...688A.106N}.

Beyond individual measurements, our sample allows for a statistical constraint on the Ly$\alpha$ escape fraction in typical star-forming galaxies at this epoch. Summing over the full sample (including both detections and non-detections), we derive a conservative upper limit on the average Ly$\alpha$ escape fraction:
\begin{equation}
\langle f_{\mathrm{esc}}^{\mathrm{Ly}\alpha} \rangle =
\frac{\sum L_{\mathrm{Ly}\alpha}}
{\sum 8.7 \times L_{\mathrm{H}\alpha}^{\mathrm{int}}},
\end{equation}
where non-detections are assumed to contribute at their $3\sigma$ upper limits and this yields $\langle f_{\rm esc}^{\rm Ly\alpha} \rangle < 0.32$. The mean Ly$\alpha$ luminosity used in this calculation is estimated via the Kaplan--Meier estimator, as detailed in Appendix~\ref{avg_lya}.

Due to the presence of censored data at the low-luminosity end, this value should be interpreted as a conservative upper bound rather than an unbiased estimate of the population average. We note that stacking analyses are limited by the small number of spectroscopic targets (six in total) and by strong sky-line residuals in the spectra, which hinder our ability to further constrain the low-luminosity regime.

In addition, the detection rate of Ly$\alpha$ emission itself provides a complementary constraint. Out of 13 H$\alpha$ emitters with MUSE coverage, only four exhibit detectable Ly$\alpha$ emission, corresponding to a Ly$\alpha$ detection fraction of $\sim$30\%. This low detection rate indicates that efficient Ly$\alpha$ escape is not a common property among H$\alpha$-bright star-forming galaxies at $z \approx 5.5$, but is instead restricted to a minority of systems with favorable ISM and CGM conditions.

\begin{deluxetable*}{lcccccccc}
\tablewidth{0pt}
\tablecaption{Catalog of H$\alpha$ emitters*\label{tab1}}
\tablehead{
\colhead{ID(a)} & \colhead{RA (J2000)} & \colhead{Dec (J2000)} & 
\colhead{redshift(b)} & \colhead{$\log (L_{\rm H\alpha}^{\rm int})$} & \colhead{$\log(L_{\rm Ly\alpha} )$} & \colhead{$f_{\rm esc}^{\rm Ly\alpha}$} & \colhead{$A_V^{\rm SED}$} & \colhead{Magnification Factor}\\
 & HH:MM:SS & DD:MM:SS & & log erg/s & log erg/s & & 
}
\startdata
17763 & 00:14:22.3 & -30:24:47.9 &  5.61432 & 42.29$\pm$0.03  & $<41.99$         & $<0.06$        & 0.1346 & 6.83  $\pm$ 1.80 \\
17939 & 00:14:20.4 & -30:24:47.1 &  5.53202 & 42.42$\pm$0.02  & $<42.13$         & $<0.06$        & 0.1669 & 3.21  $\pm$ 0.67 \\
18072 & 00:14:21.9 & -30:24:45.3 &  5.51 & 41.48$\pm$0.09  & $<42.04$         & $<0.42$        & 0.0804 & 7.28  $\pm$ 1.39 \\
20875 & 00:14:19.7 & -30:24:25.6 &  5.65540 & 41.50$\pm$0.09  & $ 42.30\pm 0.09$ & $0.71\pm 0.20$ & 0.0221 & 4.69  $\pm$ 1.01 \\
22444(c) & 00:14:18.3 & -30:24:16.1 &  5.65847 & 42.28$\pm$0.02  & $ 42.94\pm 0.02$ & $0.50\pm 0.03$ & 0.0007 & 2.76  $\pm$ 0.35 \\
27262(c) & 00:14:21.8 & -30:23:44.0 &  5.66481 & 42.89$\pm$0.01  & $ 43.42\pm 0.01$ & $0.38\pm 0.01$ & 0.0003 & 10.55 $\pm$ 3.51 \\
32191 & 00:14:25.5 & -30:23:11.9 &  5.64848 & 42.88$\pm$0.01  & $<42.04$         & $<0.02$        & 0.4971 & 1.58  $\pm$ 0.32 \\
33269 & 00:14:22.7 & -30:23:04.0 &  5.60802 & 41.86$\pm$0.04  & $ 42.41\pm 0.17$ & $0.25\pm 0.10$ & 0.0076 & 2.41  $\pm$ 0.76 \\
43154 & 00:14:08.4 & -30:22:14.8 &  5.60545 & 42.70$\pm$0.02  & $<42.45$         & $<0.06$        & 0.1416 & 3.97  $\pm$ 0.57 \\
54761 & 00:14:19.6 & -30:21:02.2 &  5.67046 & 41.88$\pm$0.05  & $<41.94$         & $<0.13$        & 0.0941 & 5.12  $\pm$ 0.54 \\
56403 & 00:14:20.1 & -30:20:51.1 &  5.65 & 41.74$\pm$0.06  & $<42.02$         & $<0.22$        & 0.0346 & 2.87  $\pm$ 0.13 \\
59930 & 00:14:21.3 & -30:20:24.0 &  5.40647 & 42.26$\pm$0.02  & $<42.27$         & $<0.11$        & 0.2026 & 1.90  $\pm$ 0.04 \\
62119 & 00:14:20.5 & -30:20:05.9 &  5.48 & 42.31$\pm$0.07  & $<42.59$         & $<0.22$        & 0.8099 & 1.73  $\pm$ 0.03 \\
\enddata
\tablecomments{* The luminosities listed in this table have not been corrected for lensing magnification. (a), We adopt the IDs from UNCOVER catalog; (b) Redshifts with 2 digits are photometric redshifts. (c) IDs 22444 and 27262 are multiple images of the same source.
After magnification correction, their intrinsic H$\alpha$ and
Ly$\alpha$ luminosities agree within 0.03~dex and 0.10~dex,
respectively. The derived $f_{\rm esc}^{\rm Ly\alpha}$ differ at
$\sim4\sigma$ (0.50$\pm$0.03 vs.\ 0.38$\pm$0.01), which may be
caused by differential lensing \citep{1995ApJ...453L..65D, 2012ApJ...761...20H}: 27262 ($\mu\approx10.55$) lies near the critical curve where compact and extended emission can be magnified by different factors, while 22444 ($\mu\approx2.76$) is in a region of gentle magnification gradient.
}
\end{deluxetable*}

\begin{figure}
    \centering
    \includegraphics[width=0.95\linewidth]{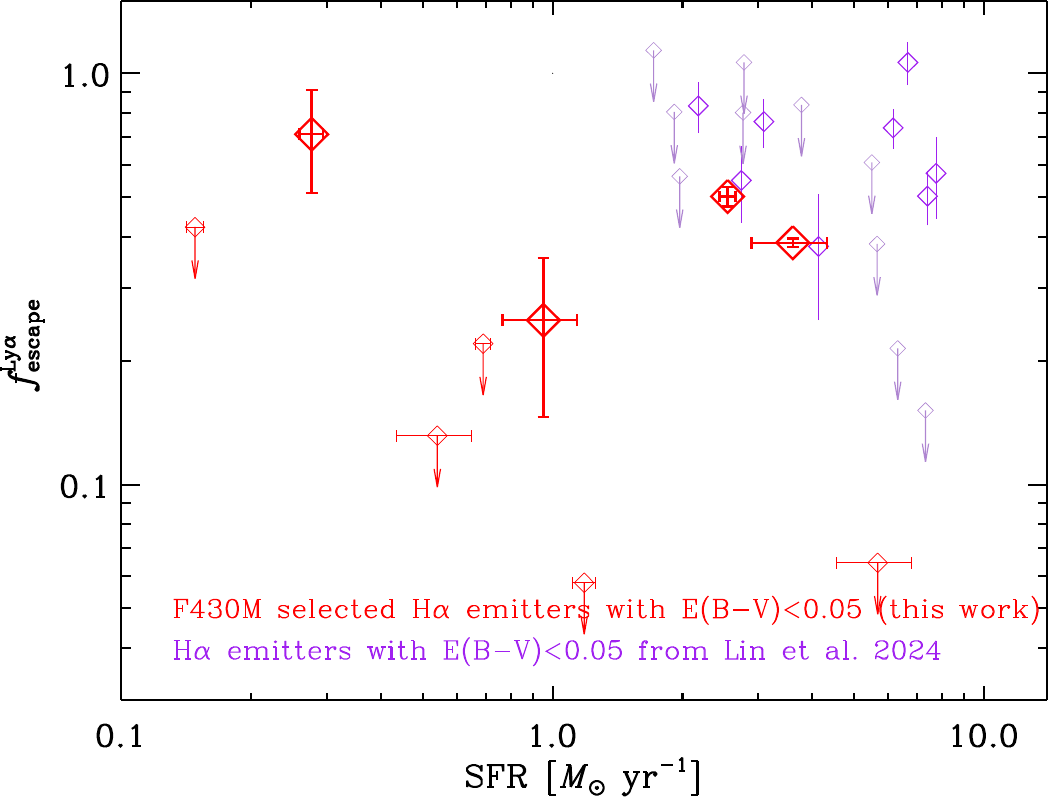}
    \caption{Ly$\alpha$ escape fraction and star formation rate distribution for galaxies with negligible dust attenuation ($E(B-V)<0.05$) in the F430M-selected H$\alpha$ emitters in this work (red diamonds), and the H$\alpha$-selected sample from \citet[][purple diamonds]{2024ApJS..272...33L}. For dust-poor systems, galaxies span a wide range of $f_{\rm esc}^{\rm Ly\alpha}$ at a given SFR, including both detections and upper limits on Ly$\alpha$ emission. The star formation rates of our sample are corrected for gravitational lensing magnification using the lenstool mass model of the A2744 cluster \citep{2022ApJ...928...87F}, allowing us to probe intrinsically lower-SFR galaxies.}
    \label{fesc_sfr}
\end{figure}

\section{Discussion: Physical Origin of Ly$\alpha$ Detection and Escape Fractions}

Among our H$\alpha$-selected galaxies at $z\simeq5.5$, Ly$\alpha$ emission is detected in only 4 out of 13 sources. The population-averaged escape fraction, including 3$\sigma$ upper limits and dust correction, is $\langle f_{\rm esc}^{\rm Ly\alpha}\rangle < 0.32$. The upper limits on $f_{\rm esc}^{\rm Ly\alpha}$ for the Ly$\alpha$-undetected sources remain low, suggesting that with deeper MUSE observations, these systems are unlikely to exhibit high Ly$\alpha$ escape fractions. This indicates that efficient Ly$\alpha$ escape occurs only in a minority of galaxies, consistent with the resonant nature of Ly$\alpha$ scattering \citep{2006A&A...460..397V, 2015PASA...32...27H}.

We examine the physical drivers of Ly$\alpha$ visibility using both the H$\alpha$ luminosities and the SED-derived dust attenuation in Figure~\ref{AvLHa}. In our H$\alpha$-selected sample, all four galaxies with detected Ly$\alpha$ emission have extremely low attenuation, with $A_V \simeq 0$, and exhibit measurable escape fractions of $f_{\rm esc}^{\rm Ly\alpha}\sim0.3$--0.5. In contrast, galaxies without Ly$\alpha$ detections span a broader range of dust attenuation, $A_V\sim0.1$--0.8. While this result alone does not imply that Ly$\alpha$ emission is confined exclusively to dust-free systems, it suggests a decline of $f_{\rm esc}^{\rm Ly\alpha}$ with increasing dust attenuation, consistent with previous results \citep[e.g.,][]{2014PASA...31...40D, 2024ApJS..272...33L}.

To further disentangle the role of dust from other physical drivers, we examine the Ly$\alpha$ escape fraction as a function of star formation rate for galaxies with negligible dust attenuation ($E(B-V)<0.05$), combining our sample with the low-reddening subsample from \citet{2024ApJS..272...33L} in Figure \ref{fesc_sfr}. 
In this dust-poor regime, Ly$\alpha$ emission is detected over a range of star formation rates of $0.2\sim 10 M_\odot\,\rm yr^{-1}$, thus the absence of Ly$\alpha$ is unlikely to be primarily due to a low production rate of ionizing photons. At a given SFR, galaxies exhibit both high and low $f_{\rm esc}^{\rm Ly\alpha}$ values, including systems with only upper limits on Ly$\alpha$ emission. Therefore, once dust attenuation is minimized, the Ly$\alpha$ escape fraction does not show a strong dependence on the instantaneous star formation rate. As shown in Figure~\ref{fesc_sfr}, the medium-band–selected H$\alpha$ emitters in the A2744 field, aided by gravitational lensing, allow us to probe intrinsically lower-SFR systems, thereby extending the dynamic range relative to H$\alpha$ samples identified via slitless spectroscopy.

\begin{figure}
    \centering
    \includegraphics[width=0.95\linewidth]{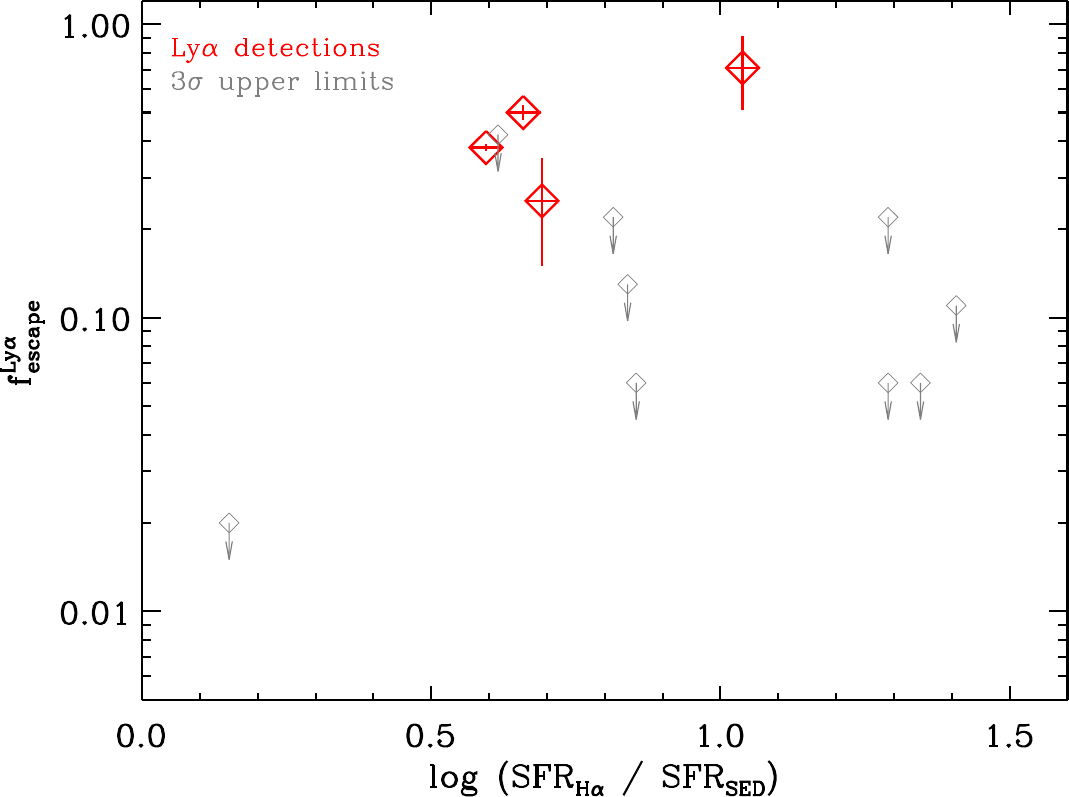}
    \caption{Ly$\alpha$ escape fraction as a function of the SFR$_{\rm H\alpha}$ and SFR$_{\rm SED}$ ratio. Red diamonds show Ly$\alpha$ detections with error bars, while gray markers with arrows denote 3$\sigma$ upper limits. Both populations span a similar range of burstiness, with no clear correlation between $f_{\rm esc}^{\rm Ly\alpha}$ and SFR$_{\rm H\alpha}$/SFR$_{\rm SED}$, supporting a stochastic picture of Ly$\alpha$ escape.}
    \label{fesc_burst}
\end{figure}

To further investigate whether recent star formation bursts facilitate Ly$\alpha$ escape, we examine the ratio of the H$\alpha$-based SFR to the SED-based SFR (SFR$_{\rm H\alpha}$/SFR$_{\rm SED}$). Figure~\ref{fesc_burst} shows $f_{\rm esc}^{\rm Ly\alpha}$ as a function of this burstiness indicator. Both Ly$\alpha$ detections and non-detections span a similar range of SFR$_{\rm H\alpha}$/SFR$_{\rm SED}$, with no clear correlation, indicating that even galaxies undergoing recent starbursts do not necessarily exhibit high Ly$\alpha$ escape fractions.

\begin{figure}
    \centering
    \includegraphics[width=0.95\linewidth]{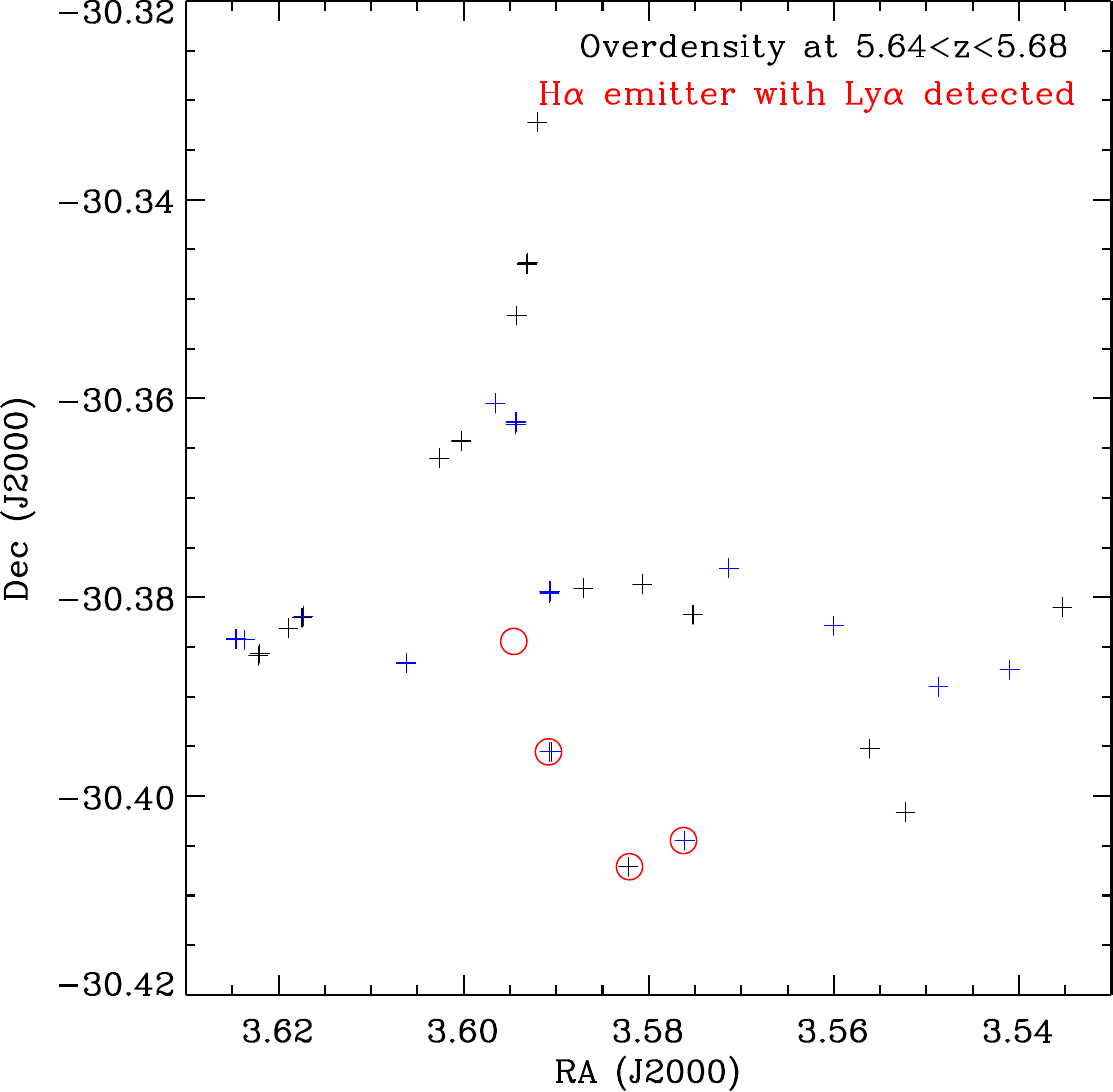}
    \caption{Spatial distribution of the galaxy overdensity at $5.64<z<5.68$ behind Abell~2744. Crosses mark spectroscopically confirmed sources in this redshift range from the ALT catalog \citep{2024arXiv241001874N}. Objects with ${\rm N\_LINES\_DETECTED}>1$, indicating more secure spectroscopic redshifts, are highlighted in blue. Ly$\alpha$-detected galaxies from our H$\alpha$-selected sample are shown as red circles. Three of the four Ly$\alpha$ detections lie within the overdense region, while the remaining source is projected close to it, suggesting a possible environmental effect. The physical size of the field shown is $\sim$2~Mpc at $z=5.66$.
    }
    \label{clustering}
\end{figure}

Interestingly, previous studies have identified a galaxy overdensity at
$z\simeq5.64$--5.68 in this field \citep{2025ApJ...982..153M}. 
Three of our Ly$\alpha$-detected sources (of which 22444 and 27262 are multiple images) lie within this narrow redshift range. We show the spatial distribution of this overdensity from ALT catalog in Figure \ref{clustering}. The three Ly$\alpha$ detected targets are found to be closely clustered in projected space. This spatial association, albeit based on small number statistics 
tentatively suggests that local environment may play a role in facilitating Ly$\alpha$ escape.

In overdense regions, enhanced ionizing radiation from neighboring galaxies may reduce the neutral hydrogen fraction in the surrounding circumgalactic and intergalactic medium, effectively increasing the transparency to Ly$\alpha$ photons. Such enhanced transmission, potentially linked to the emergence of large-scale ionized regions (or ``ionized bubbles'') during the late stages of reionization, has been proposed as a mechanism to boost Ly$\alpha$ visibility. However, recent studies paint a more nuanced picture of this process, revealing that the correlation between galaxy density and IGM transmission is highly scale-dependent and rapidly evolving. Observations at $z \sim 6$ indicate that while excess IGM transmission is detected on large scales of $\sim 10-60$ cMpc around galaxies \citep{2020MNRAS.494.1560M, 2025arXiv250307074K, 2025MNRAS.542.1952M, 2026arXiv260502539H}, absorption from local neutral gas overdensities often dominates at smaller scales ($\lesssim 10$ cMpc) \citep{2025arXiv250307074K, 2026ApJ...997..280K}. Furthermore, \citet{2026ApJ...997..280K} demonstrate that this relationship flips with redshift—from suppression in overdense regions at $z < 5.5$ to enhancement at $z > 5.7$—suggesting a transition from local ionization dominance to a rising background radiation field. Given the limited statistical significance of our current sample size, we cannot definitively rule out that the observed Ly$\alpha$ fractions are subject to cosmic variance. Nevertheless, the apparent concentration of Ly$\alpha$ detections in this overdense region at $z \sim 5.66$ (a redshift that falls precisely within the transitional regime identified by \citet{2026ApJ...997..280K}) is broadly consistent with a scenario in which environmental effects modulate Ly$\alpha$ escape. Rather than purely intrinsic production, the early stages of large-scale bubble percolation may begin to alter external attenuation at these intermediate scales and redshifts.

Taken together, these results suggest that Ly$\alpha$ escape depends on a combination of internal ISM conditions (low dust content, low neutral gas covering fraction) and external factors, including scattering in the CGM, attenuation by the IGM, and the local galaxy environment. The association of 3 (out of 4) Ly$\alpha$ detections (corresponding to 3 unique systems; see Section 2.2) with a compact overdensity at $z\simeq5.66$ suggests that favorable environments may enhance Ly$\alpha$ visibility by reducing circumgalactic opacity or providing ionized channels for photon escape. The Ly$\alpha$-detected sources may therefore trace systems observed along favorable sightlines or residing in environments conducive to Ly$\alpha$ transmission, whereas the non-detections may reflect the combined effects of internal absorption and external suppression.

\section{Summary}

We construct a sample of 13 H$\alpha$-selected galaxies at $z\simeq5.5$ with intrinsic star formation rates of SFR $\gtrsim 0.1\,M_\odot\,{\rm yr^{-1}}$, identified using JWST/NIRCam F430M imaging and covered by MUSE spectroscopy. By selecting galaxies via H$\alpha$ rather than Ly$\alpha$, this sample provides a direct and Ly$\alpha$-unbiased probe of the Ly$\alpha$ escape process, largely independent of the stochastic visibility of Ly$\alpha$ emission. Four galaxies show detectable Ly$\alpha$ emission, while the remainder are undetected down to the MUSE sensitivity limits, highlighting the diverse and stochastic nature of Ly$\alpha$ escape in typical star-forming galaxies at this epoch. Notably, three of the four Ly$\alpha$ detections are associated with a known overdense structure at $z\simeq5.66$, suggesting that environmental factors may further facilitate Ly$\alpha$ escape.

This H$\alpha$-selected framework provides a robust and scalable method for studying Ly$\alpha$ escape across cosmic time by combining JWST medium- or narrow-band imaging with ground-based spectroscopic follow-up (e.g., VLT/MUSE), offering a general and less biased approach that is not restricted to a specific redshift or dataset. In future work, we will extend this method to multiple redshifts using JWST/NIRCam medium-band imaging (e.g., F360M and F480M targeting H$\alpha$ at $z\sim4.4$ and $z\sim6.3$, respectively), enabling uniform and less biased constraints on the evolution of the Ly$\alpha$ escape fraction across cosmic time by constructing SFR-limited samples at different redshifts.

\begin{acknowledgments}
We would like to thank the referee for their careful reading and helpful comments.
This work is supported by the National Key R\&D Program of China No.2025YFF0510603, the National Natural Science Foundation of China (grant 12373009), the CAS Project for Young Scientists in Basic Research Grant No. YSBR-062, the China Manned Space Program with grant no. CMS-CSST-2025-A06, and the Fundamental Research Funds for the Central Universities. XW acknowledges the support by the Xiaomi Young Talents Program, and the work carried out, in part, at the Swinburne University of Technology, sponsored by the ACAMAR visiting fellowship. ZYZ acknowledges the supports by the Shanghai Leading Talent Program of Eastern Talent Plan (LJ2025051) and the China-Chile Joint Research Fund (CCJRF No. 1906). This work is supported by National SKA Program of China No. 2025SKA0150101. This work is sponsored (in part) by the Chinese Academy of Sciences (CAS) through a grant to the CAS South America Center for Astronomy. This work is supported by the China Manned Space Program with grant no. CMS-CSST-2025-A07. C.C. is supported by Chinese Academy of Sciences South America Center for Astronomy (CASSACA) Key Research Project E52H540101 and E52H540301. E.I. and J.M. gratefully acknowledge financial support from ANID - MILENIO - NCN2024\_112. 

(Some of) The data products presented herein were retrieved from the Dawn JWST Archive (DJA). DJA is an initiative of the Cosmic Dawn Center (DAWN), which is funded by the Danish National Research Foundation under grant DNRF140.

This work utilizes gravitational lensing models produced by PIs Brada\v{c}, Natarajan \& Kneib (CATS), Merten \& Zitrin, Sharon, Williams, Keeton, Bernstein and Diego, and the GLAFIC group. This lens modeling was partially funded by the HST Frontier Fields program conducted by STScI. STScI is operated by the Association of Universities for Research in Astronomy, Inc. under NASA contract NAS 5-26555. The lens models were obtained from the Mikulski Archive for Space Telescopes (MAST).

All the HST and JWST data used in this paper can be found in MAST: \dataset[https://doi:10.17909/1esp-hh29]{https://doi:10.17909/1esp-hh29}. We would like to thank the MAGNIF project for providing valuable insights that inspired part of this work.

We acknowledge the use of ChatGPT (OpenAI) for assistance with English language editing.

\end{acknowledgments}

\begin{contribution}


Dr. Zhenya Zheng and Dr. Cheng Cheng came up with the initial research concept and Cheng Cheng prepared the manuscript. Dr. Cheng Cheng prepared the first draft of the manuscript. All authors contributed to the analysis and interpretation of the results and to the writing of the manuscript.

\end{contribution}

%
\facilities{HST, JWST}

\software{astropy \citep{2013A&A...558A..33A,2018AJ....156..123A,2022ApJ...935..167A},  
          Source Extractor \citep{1996A&AS..117..393B}
          }


\appendix

\section{H$\alpha$ Flux Estimation from F430M Excess}\label{Haflux}

As an independent check, we estimate the H$\alpha$ flux using the F430M excess relative to the adjacent F444W band. Assuming that the continuum can be approximated by the F444W flux, the H$\alpha$ line flux can be estimated from the F430M excess as
\begin{equation}\label{f430f444}
    F_{\rm H\alpha} = \Delta {\rm F430M} \frac{f_{\rm F430M} - f_{\rm F444W}}{1-\Delta {\rm F430M}/\Delta {\rm F444W}},
\end{equation}
where $f_{\rm F430M}$ is the observed flux density in F430M, $f_{\rm F444W}$ is the continuum flux density at the F430M wavelength estimated from F444W flux, and $\Delta{\rm F430M} = 2315.31$\AA\ and $\Delta{\rm F444W} = 11144.05$\AA\ are the width of the F430M and F444W filters. The denominator term corrects for the contribution of the emission line to the continuum estimate. Figure~\ref{LHa} compares the H$\alpha$ luminosities derived from this method with those obtained from the spectroscopic measurements. The two methods yield consistent H$\alpha$ luminosities within the measurement uncertainties, demonstrating the robustness of our H$\alpha$ flux estimates.

\setcounter{figure}{0}
\renewcommand{\thefigure}{A-\arabic{figure}}

\begin{figure}
    \centering
    \includegraphics[width=0.5\linewidth]{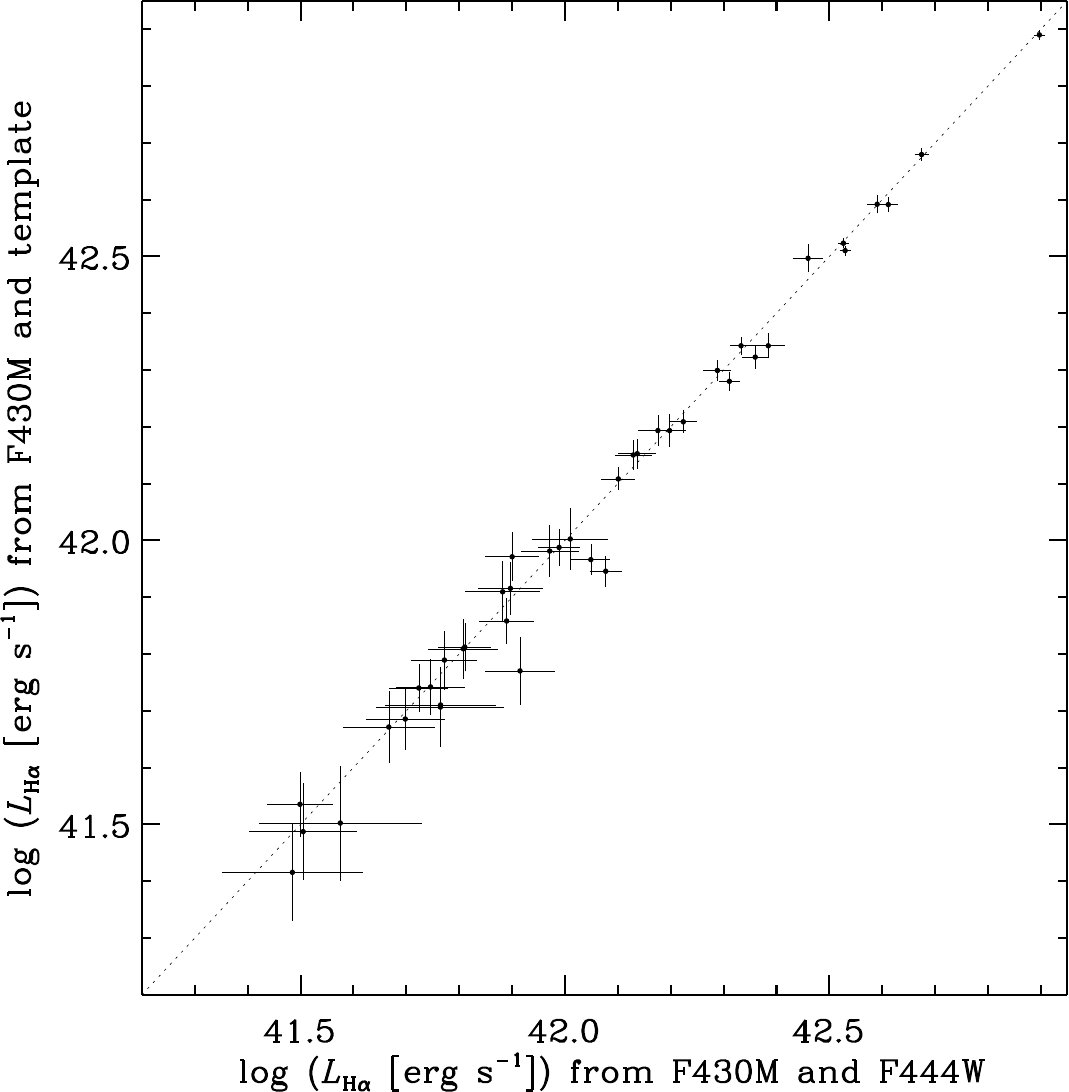}
    \caption{Comparison of the H$\alpha$ luminosities derived using different continuum
estimation methods for the F430M band.
The two measurements are consistent within the quoted uncertainties.}
    \label{LHa}
\end{figure}

\section{Average Lyman-$\alpha$ luminosity including upper limits}\label{avg_lya}

The mean Ly$\alpha$ luminosity is estimated using the Kaplan--Meier (KM) estimator \citep{0a032d9f-e977-3fbb-ab72-070bdac732b0},
treating non-detections as upper limits \citep{1985ApJ...293..192F, 1985ApJ...293..178S}. The KM estimator is a non-parametric method widely used to handle censored data, and its survival function $S(L)$ represents the probability that the Ly$\alpha$ luminosity exceeds $L$, 
\[
S(L) = P(L_{\rm Ly\alpha} > L).
\]
The population-averaged luminosity is obtained by integrating the KM survival function,
\[
\langle L_{\mathrm{Ly}\alpha} \rangle = \int_0^{\infty} S(L)\, dL.
\]
This approach naturally incorporates both Ly$\alpha$detections and non-detections, providing a conservative estimate of the typical Ly$\alpha$escape fraction for the H$\alpha$-selected population.




\end{document}